\documentclass[nopreprintline,5p,times,twocolumn]{elsarticle}

\usepackage{amsmath,amssymb,amsfonts}
\usepackage{algorithmic}
\usepackage{graphicx}
\usepackage[normalem]{ulem}

\usepackage{xcolor}
\usepackage{siunitx}
\usepackage{subcaption} 
\usepackage{algorithm}

\newcommand{\bigO}{\mathcal{O}}
\DeclareBinaryPrefix\Kilo{K}{10}
\DeclareSIUnit\transfer{T}
\newcommand*{\etal}{\textit{et al.~}}
\usepackage[percent]{overpic}

\usepackage{colortbl}
\usepackage{xcolor}
\usepackage{multirow}

\usepackage{etoolbox}

\usepackage{booktabs} 
\usepackage{url}

\usepackage{eso-pic}
\usepackage{hyperref}

\journal{Journal of Parallel and Distributed Computing }

\begin{document}

\begin{frontmatter}

\title{Intelligent colocation of HPC workloads\tnoteref{t1,t2,t3}}
\tnotetext[t1]{This article is an extension of our previous work~(Zacarias \etal 2019) \cite{Zacarias2019colocation}, which was presented at the {International Symposium on Computer Architecture and High Performance Computing} ({SBAC-PAD 2019}).
}
\tnotetext[t2]{\copyright~2021. This manuscript version is made available under the CC-BY-NC-ND 4.0 license \href{http://creativecommons.org/licenses/by-nc-nd/4.0/}{http://creativecommons.org/licenses/by-nc-nd/4.0/}
}
\tnotetext[t3]{
For the final article published in Elsevier's Journal of Parallel and Distributed Computing refer to
\href{https://doi.org/10.1016/j.jpdc.2021.02.010}{https://doi.org/10.1016/j.jpdc.2021.02.010}
}

\author[l1,l2,l3]{Felippe Vieira Zacarias}
\author[l1,l5]{Vinicius Petrucci}
\author[l4]{Rajiv Nishtala}
\author[l3]{Paul Carpenter}
\author[l5]{Daniel Moss\'e}

\address[l1]{Universidade Federal da Bahia}
\address[l2]{Universitat Polit\`ecnica de Catalunya}
\address[l3]{Barcelona Supercomputing Center}
\address[l4]{Coop, Norway/Norwegian University of Science and Technology, Norway}
\address[l5]{University of Pittsburgh}

\begin{abstract}
Many HPC applications 
suffer from a bottleneck
in the shared caches, instruction execution units, I/O or memory bandwidth, even though the remaining resources may be underutilized. It is hard for developers and runtime systems to ensure that all critical resources are fully exploited by a single application, so an attractive technique for increasing HPC system utilization is to colocate multiple applications on the same server. When applications share critical resources, however, contention on shared resources may lead to reduced application performance.

In this paper, we show that server efficiency can be improved by first modeling the expected performance degradation of colocated applications based on measured hardware performance counters, and then exploiting the model to determine an optimized mix of colocated applications. This paper presents a new intelligent resource manager and makes the following contributions: (1) a new machine learning model to predict the performance degradation of colocated applications based on hardware counters and (2) an intelligent scheduling scheme deployed on an existing resource manager to enable application co-scheduling with minimum performance degradation. Our results show that our approach achieves performance improvements of \SI{7}{\%} (avg) and \SI{12}{\%} (max) compared to the standard policy commonly used by existing job managers.

\end{abstract}


\end{frontmatter}

\section{Introduction}

Data center providers need to maximize server utilization, in order to obtain the greatest possible benefit from their large capital investments~\cite{mars2011bubble,yang2013bubble}. Many HPC applications, however, only
achieve a fraction of the theoretical peak performance, even when they have been carefully optimized~\cite{breitbart2015case}.
This can lead to a substantial waste of resources across the whole data center.

In HPC systems, resource efficiency is an important and growing concern to achieving exascale computing performance.
To reach exascale  using current  technology, would require an unrealistic amount of energy. 
Even worse, the electricity bill to sustain these platforms considering their lifespan can be roughly equal to their hardware cost~\cite{dutot2017towards}. While energy-proportional designs~\cite{barroso2007case} could be
a solution for HPC systems, this technology is still maturing. Thus, exascale systems are expected to be resource-constrained in the near future, which means the amount of provisioned power will severely limit the
scalability to meet new user demands~\cite{gholkar2016power,patki2013exploring}.

Under a resource-constrained server environment, minimizing resource usage while meeting performance requirements is key to keeping up with increased computational demands. Techniques like hardware over-provisioning can be
applied as a solution for systems with strict power bounds. The idea behind over-provisioning is to use less power per node and thereby allowing more nodes in the system~\cite{gholkar2016power}. In real settings,
over-provisioning can be implemented by enforcing socket-level power limits with Intel's RAPL technology~\cite{guide2011intel}. RAPL relies on updating registers to manage power usage of the server components (processor,
DRAM, GPUs, etc.). 
It works by monitoring low-level hardware events to estimate power consumption~\cite{zhang2016maximizing}, and it adapts the processor voltage and frequency to meet the desired power cap during a specified time interval. 

Techniques like DVFS also adapt the processor voltage and frequency to reduce processor power consumption. Lower frequencies require less power, potentially resulting in energy reduction in the
system~\cite{rountree2009adagio}. Although this can improve energy efficiency, it may negatively impact the processor performance. 
Either DVFS or RAPL alone is insufficient for running in an over-provisioned environment, since it only enforces power bound for individual components, such as the CPU. 
Then, the power bound across all components needs to be enforced by a global scheduler to avoid violating the system bound~\cite{ellsworth2015dynamic}.
A promising way to increase overall system utilization and efficiency is to run multiple applications concurrently on a server node, an approach that is known as workload
colocation~\cite{mars2011bubble,yang2013bubble,Nishtala2013emsoft,tang2013reqos,dwyer2012practical,Zacarias2019colocation}. The biggest disadvantage of workload colocation is the potential degradation in application performance due to sharing of
resources such as caches, memory controllers, data prefetchers, and I/O devices. Such degradation is hard to predict in real systems, and it is impractical to measure the degradation of all pairs of applications ahead of
time. 

Due to the uncertain degradation effects, HPC systems usually do not support the sharing of resources in the same computing node among applications~\cite{de2017disallowing,mars2011bubble}. Nevertheless, workload
colocation does have a substantial potential to improve system throughput, especially when the colocated applications are bottlenecked on different resources~\cite{breitbart2015case,Zacarias2019colocation}. Note that this improvement in system
utilization is made without any need to modify the application's source code. 

Machine learning techniques have the potential to find a more complex relationship among input features (in our case, hardware \textit{performance monitoring counters}, or PMCs) and the prediction of (in our case) the slowdown experienced by the colocated applications. Given the diverse nature of applications in HPC systems, machine learning models have the capability to generalize for new different applications. 
Although (some) machine learning  approaches are computationally intensive, we observe, in our experiments, that they can be used with acceptable execution overhead with some optimizations (e.g.,~reducing the number of trees used in a random forest model)

This paper presents an approach that encompasses a machine learning model and its integration into an intelligent application scheduler. 
The goal is to reduce the \emph{makespan}, which is the time between the start of the first application and the end of the last application in a job queue of fixed size.
We make the following contributions:
\begin{enumerate}
	
	\item We design and build prediction models to estimate performance degradation due to workload colocation.
	Our best model (Random Forest) is capable of providing good accuracy (81\%) in predicting performance degradation.

	\item  We present a workload management solution based on machine learning for scheduling jobs, leveraging a given predicted degradation between colocated applications in the HPC system.
	\item We evaluate our solution through an actual implementation deployed on an existing resource manager executing multithreaded applications from PARSEC, NPB, Splash, and Rodinia. We show a reduction in makespan (7\% on average and 12\% maximum) over the existing resource manager.

\end{enumerate}

The key advantage of our technique is that it 
does not require application modifications or annotations; rather it is based on PMC data for deciding which applications can be run together with minimum degradation in a particular server.

\section{Motivation}
\label{sec:motivation}

We motivate our work by discussing the challenges posed by resource colocation and how machine learning can be a favorable strategy  to solve the server efficiency problem. We also conduct an experiment to show that prior work based on heuristics may fail to solve this problem.

\subsection{Resource Colocation}

Multi-core processors share both on and off-chip resources, such as caches, interconnects, and memory controllers~\cite{zhuravlev2012survey}. It is known that applications running on different cores will experience
contention due to these shared resources. Having a methodology that is capable of predicting how well a system will run in a particular colocation scenario would be very useful, as without prediction, profiling all possible
colocations' performance beforehand to guide scheduling decisions can be prohibitively expensive. 
Using a machine learning system can be very useful, as we train on a subset of the applications and have a model that could generalize to the other new application colocations.

A large portion of previous research on performance prediction on multi-core is focused on contention for shared last level caches~\cite{rai2010performance}. Prior work has attempted to estimate the performance slowdown of applications due to interference on shared resources~\cite{subramanian2015application}. They focus on developing techniques to predict the application behavior using indicators of cache usage. This behavior can either predict the extent to which an application suffers under colocation or the extent to which it can degrade its colocated application. These studies resulted in classification schemes for understanding the performance slowdown faced by applications~\cite{rai2010performance}.

Furthermore, many server applications can achieve only a fraction of the theoretical peak performance, even when carefully optimized to get close to the system's limit~\cite{breitbart2015case}. The sub-utilization tends to increase cores idle time, thus resulting in an over-provisioning that negatively impacts the utilization of the entire system~\cite{mars2011bubble}.

\begin{figure}[t]
	\begin{center}

		\includegraphics[scale=0.15]{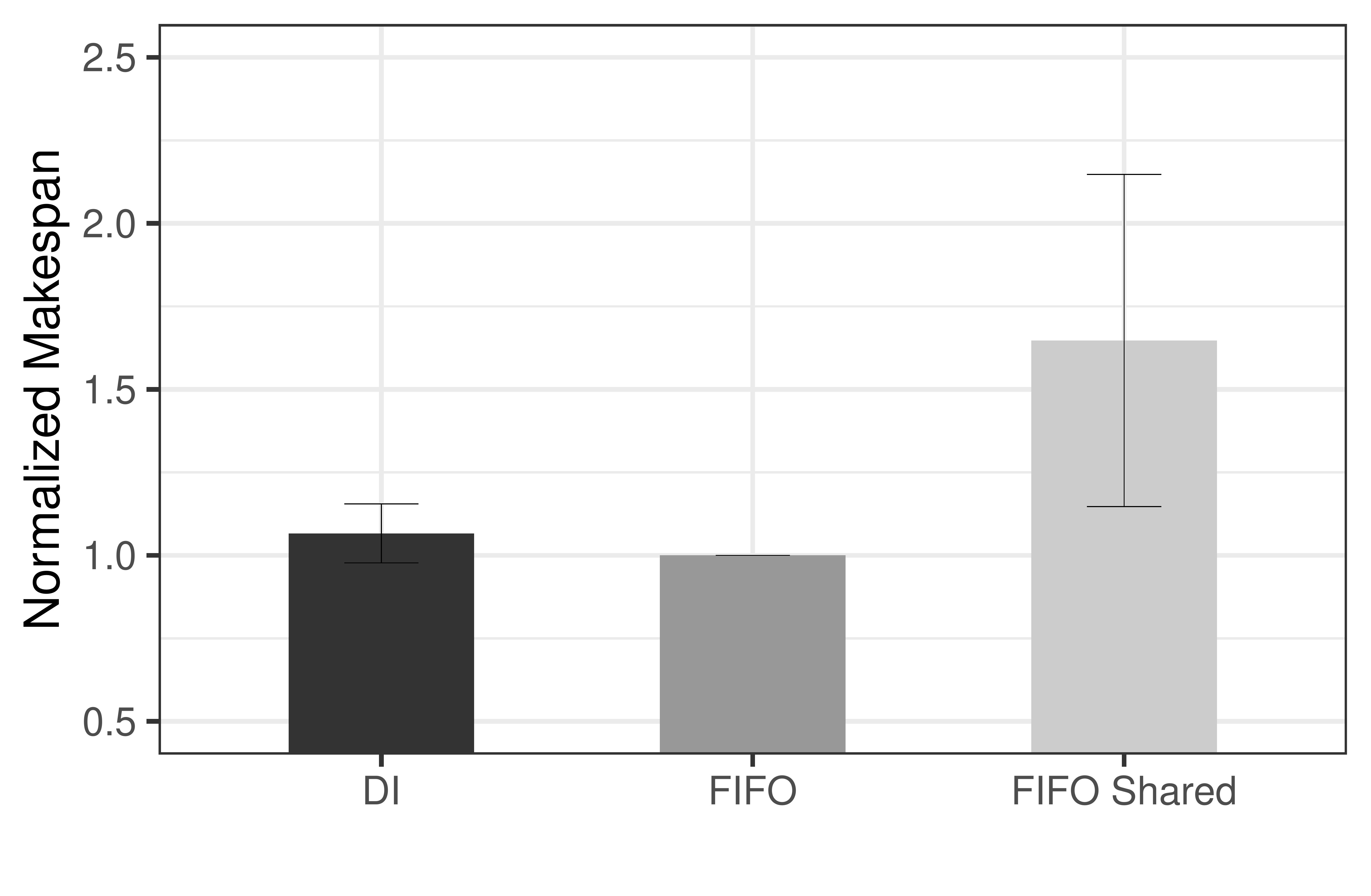}
		\caption{Normalized makespan for Distributed Intensity~\cite{blagodurov2010contention} (prior work) vs FIFO sequential (no-sharing) and FIFO shared policies, while executing 20 randomized application queues (see Section \ref{sec:results} for details on experimental setup).}
		\label{fig:di_x_fifo}
	\end{center}
\end{figure}

\begin{figure*}[bht]
	\begin{center}

		\includegraphics[scale=0.455]{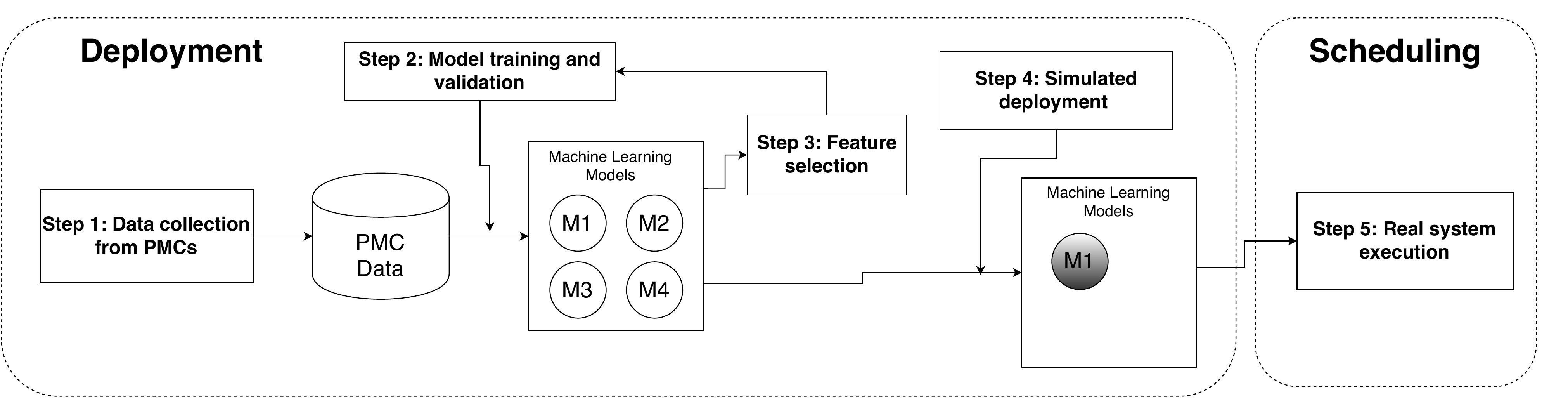}
		\caption{Overview of our approach consisting of two major phases: (a) offline deployment of a machine learning model for performance degradation prediction due to colocation and (b) online scheduling of the submitted job queue given a machine learning model. At the deployment stage, PMC data is collected, given to machine learning models for training and validation (with an optional step of feature selection/simplification), and models (M1--M4 in the figure) are evaluated using offline simulation. The best model (M1 in the figure) is used for scheduling in the real system deployment.}

		\label{fig:overall_scheme}
	\end{center}
\end{figure*}

\subsection{Heuristic-based Solutions}

Prior research has addressed the workload colocation problem~\cite{blagodurov2010contention,merkel2010resource,mishraesp} (see Section \ref{sec:related}), which favors co-scheduling applications with high variance in the shared resource (e.g., the Last-Level Cache, or LLC). Distributed Intensity (DI)~\cite{blagodurov2010contention} is an example of such a heuristic, which collects the LLC miss rate via hardware counters and avoids co-scheduling two applications with high LLC miss rates in the same memory hierarchy. By exploring workload colocation, we can minimize the total time to execute all applications, i.e. the \emph{makespan}.

Figure~\ref{fig:di_x_fifo} shows the result of an experiment in which DI actually increases the makespan by 7\% compared to the default FIFO policy in a typical job manager system (see Section \ref{sec:results} for details), but could improve by 30\% over FIFO shared (allowing applications to run on the same node). This behavior is due to ``bad'' pairs co-scheduled by DI, which is observed in most executions. In other words, relying on simple heuristics (for example, using cache misses as a sole indicator) was not sufficient to minimize overall resource contention and improve server efficiency execution.

In contrast to prior heuristics, the key idea of our approach is to build a model that abstracts the application, by predicting its degradation introduced by colocating a pair of applications through automatic analysis of hardware \textit{performance monitoring counters} (PMCs).
Each of the PMCs tracks the occurrence of a single type of event (single hardware resource) with negligible overhead, but they do not directly reveal how much degradation the application will suffer when it is colocated with another application. In addition, there are many PMCs that can be used for monitoring; only a subset of them can be collected at a time, and only a (a different) subset is necessary and sufficient to reveal the degradation due to colocation. It is hard to intuitively identify the right PMCs and manually build the right model that maps the PMCs to the predicted degradation. For these reasons, we build and evaluate machine learning models to help us solve the problem. 

\section{Intelligent Workload Management}\label{sec:approach}

This section describes our solution based on PMCs to perform intelligent colocation of workloads in HPC systems.

\subsection{Overview}

Figure~\ref{fig:overall_scheme} presents a high-level overview of our solution, organized in an offline phase (Deployment), and an online phase (Scheduling). The deployment phase is responsible for training a machine learning model to estimate performance degradation due to workload colocation. The scheduling phase takes as input a job queue to execute in the server and decides on optimized colocations and a new ordering for those jobs.

A typical cluster scheduler in a workload manager uses a flavor of bin packing algorithm to assign ready jobs to physical nodes based on user-defined resource demands and constraints. In our solution, we keep the same distribution methodology performed by an existing workload manager and add to it non-exclusive access to physical nodes in a degradation-aware fashion. We design our solution to be readily deployed on an existing workload manager.

The machine learning model built in the deployment phase is used to predict the degradation for any pair of applications submitted to execute in the queue.
We focus on finding a good colocation for pairs of applications as in~\cite{mars2011bubble,yang2013bubble}. We use the maximum runtime execution within each pair to characterize degradation. We assume that each application has executed once in the system and its profile (that is, the PMCs measured for the application execution alone on the target system) is accessible to the workload manager.

In the scheduling phase, our technique receives as input both the list of jobs in the ready queue and the profile data for such jobs. Next, it uses the trained model to predict the degradation of two particular jobs if they were to run colocated with each other. Finally, it generates an optimized schedule queue that minimizes the overall runtime between the pairs of applications. Note that a new job queue is not created; the job queue provides the metadata for the workload manager to execute the pair in the right order.

To build the solution we follow these steps (as depicted in Figure~\ref{fig:overall_scheme}):

\begin{itemize}
	\item Step 1: Data collection from PMCs (Section \ref{sec:data-collect})
	\item Step 2: Model training and validation (Section \ref{sec:model-train})
	\item Step 3: Input feature selection (Section \ref{sec:feat-sel})
	\item Step 4: Simulated deployment (Section \ref{sec:sim-deploy})
	\item Step 5: Real system deployment (Section \ref{sec:real-deploy})
\end{itemize}

Below we detail each step of our solution:

\subsection{PMC Dataset Collection}\label{sec:data-collect}

We collect PMCs and execution time from the applications to build a training set, which is usual in many data centers~\cite{Kanev2015wsc}. An application profile (input features) consists of its execution time on the target system and its collected PMCs. 

We create the input dataset used for training the models, as follows. First, we execute each application alone in the system (without any other user applications).
This was necessary to obtain a baseline for the application execution time and hardware counters considering the ideal 
condition: the application has all resources available without competing with other applications. We allow each application to use all available cores. All applications were executed multiple times and we collect the mean, minimum, maximum value, and the standard deviation for all of the PMCs shown in Table~\ref{tab:table_metrics}.

The baseline execution time is also needed to calculate the degradation experienced by application $i$ during colocation with application $j$ (Equation~\ref{eq:slowdown}), which is calculated as the percentage increase in application $i$'s execution time, where  $T_{alone_i}$ is under the ideal conditions and $T_{coloc_{ij}}$ when applications $i$ and $j$ execute colocated with each other.
\begin{equation}\label{eq:slowdown}
	Deg_{ij} = 100 \times (T_{\mathrm{coloc}_{ij}} - T_{\mathrm{alone}_{i}})/T_{\mathrm{alone}_{i}}
\end{equation}

After collecting $T_{alone_i}$ for all $i$, we execute every (primary) application concurrently with another (interfering) application. We execute all possible pairs of applications used for training by starting the two applications at the same time. If the interfering application finishes before the primary application, we restart it to keep the primary application under contention during its entire execution. 

Although the $O(n^2)$ complexity (for $n$ applications) for data collecting from job colocation can be high if $n$ is large, in the long term, this cost can be reduced by executing a subset of the pairs of common applications, and estimating the rest (e.g., via matrix factorization), or by using fast $O(n)$ microbenchmarking as in Bubble-Up~\cite{mars2011bubble}. Note that the $O(n^2)$ complexity is dependent on the number of applications used for training, and not the total number of applications deployed on the system.

The resulting training dataset contains three parts: (a) the PMC data for the primary application when running alone on the system; (b) the PMC data for the interfering application running alone on the system and (c) the performance degradation suffered by the primary application for this pair. 
Prior to the training of our models we identify applications that are not affected at all by running in shared mode and consequently might introduce small negative degradation values as noises to the input file. We then alter to 0 the negative values that appear to represent no apparent degradation in the colocated application. 

\begin{table}[t]
	\centering
	\caption{Hardware counters and derived metrics.}
    
	\label{tab:table_metrics}
    \begin{tabular}
    {
    p{1.3cm}>{\raggedright\let\newline\\\arraybackslash\hspace{0pt}}p{6.5cm}}
    \toprule
    
    Events & Counters/PMCs \\
    \midrule
	
    \multirow[t]{3}{*}{Hardware}  & 	cycles, instructions, resource\_stalls.any,    \\
                       & branch\_instructions, stalled\_cycles\_frontend, \\ 
                       &  stalled\_cycles\_backend,branch\_misses \\[1mm]
    \multirow[t]{1}{*}{Software}  	  & page\_faults,  context\_switches, cpu\_migrations \\[1mm]
     \multirow[t]{9}{*}{Cache}  	  & cache\_references, cache\_misses,\\
     							  & LLC\_prefetches, LLC\_prefetch\_misses, l2\_rqsts.demand\_data\_rd\_hit\\                                  
                                  & l2\_rqsts.pf\_hit, l2\_l1d\_wb\_rqsts.miss   \\ 
                                  &	l2\_lines\_out.pf\_clean, l2\_lines\_out.pf\_dirty\\
                                  & l2\_rqsts.all\_pf,  l1d.allocated\_in\_m\\ 
                                  &	l2\_lines\_out.demand\_clean, l1d.eviction\\ 
                                  &	l2\_rqsts.all\_demand\_data\_rd,  l2\_lines\_in.all, L1\_dcache\_store\_misses,\\ 
                                  &	L1\_dcache\_load\_misses, L1\_dcache\_loads\\ 
                                  &	L1\_dcache\_prefetch\_misses, l1d.replacement\\[1mm]
    \multirow[t]{4}{*}{Memory}  & mem\_uops\_retired.all\_stores \\
    								& mem\_uops\_retired.all\_loads\\ 
                                    & mem\_load\_uops\_retired.llc\_miss\\
                                    & mem\_load\_uops\_retired.llc\_hit\\[1mm]

     \multirow[t]{2}{*}{\emph{Calculated}}  & \emph{IPC, cache\_ref\_per\_instructions, CPU\_usage} \\
                                & \emph{cache\_misses\_per\_instructions,  miss\_ratio}\\ 
	\bottomrule
    \end{tabular}
    
\end{table}

\subsection{Model Training and Validation}\label{sec:model-train}

\begin{figure}[ht]
	\begin{center}
		\includegraphics[scale=0.6]{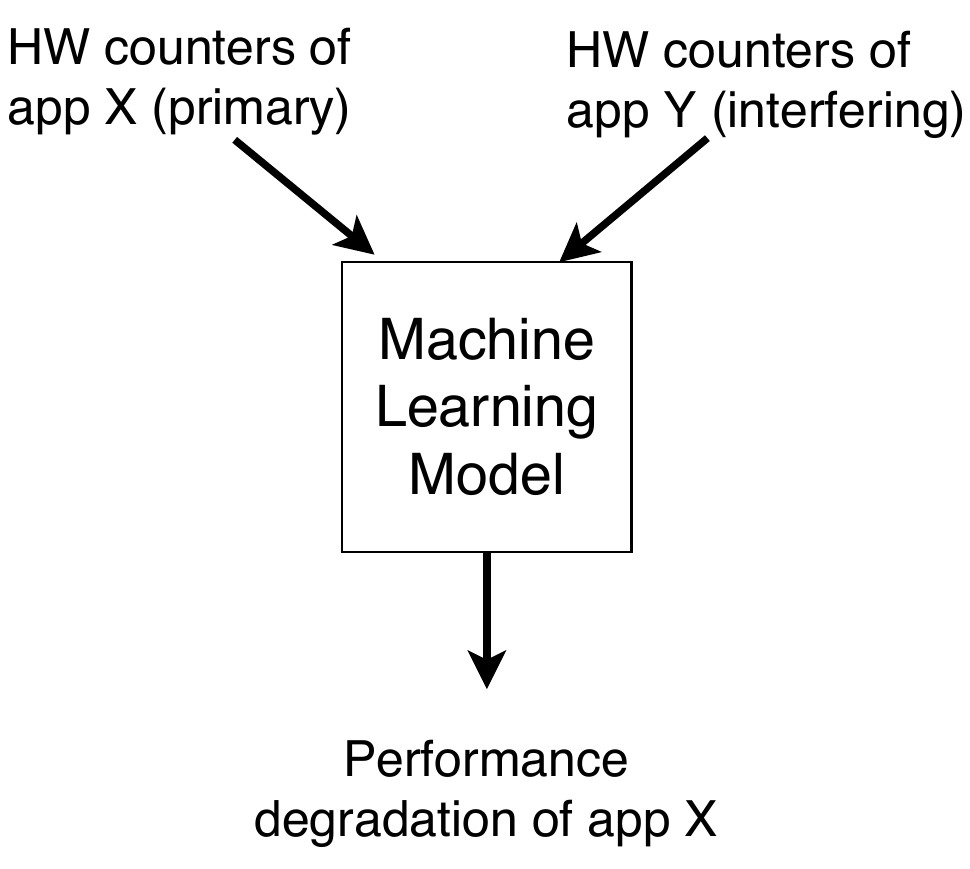}
		\caption{Overview of how the prediction model works for estimating the degradation of a pair of applications.}
		\label{fig:model}
	\end{center}
\end{figure}

Our methodology aims to estimate the degradation of two applications running concurrently on the same machine based 
on the PMC data from these applications. As illustrated in  
Figure~\ref{fig:model}, given the PMC input data for an application $X$ (the primary one) and application $Y$ (the interfering one), the trained model will output the expected degradation suffered by the primary application $X$ when colocated with application $Y$. 

We explore and evaluate the following well-known machine learning models in our work. \textit{Elastic Net}~\cite{Zou05regularizationand}, a regularization method that does variable selection and shrinkage of regression coefficients applying penalty, and it groups  strongly correlated variables; \textit{Support Vector Machine (SVM) Regressor}~\cite{Boser:1992:TAO:130385.130401}, a generalization of SVM to solve regression problems; \textit{Random Forest Regressor}~\cite{Breiman:2001:RF:570181.570182}, a supervised learning algorithm that assembles multiple decision trees to predict a decision for a given mapping function; and  
\textit{Multilayer Perceptron (MLP) Regressor}~\cite{hinton1990connectionist} that learns a non-linear function allowing one or more hidden layers between the input and output layers. These layers work by specifying a set of ``neurons'' that can propagate the inputs through a network using a series of functions located at each node.  

\subsubsection{Training Phase}\label{subsec:r2}

From the complete dataset, we hold 30\% of the data (randomly chosen) to validate and test the final model.  For the training and validation of the described models, we iterated over them applying 5-fold cross-validation to achieve higher scores during the training phase. The method splits the other 70\% of the input data into \emph{k} number of subsets (in our case \emph{k $=$ 5}), then performs the training on \emph{k-1} subsets, leaving one subset for the evaluation of the trained model and iterating \emph{k} times with a different subset for testing each time. The same process is repeated \emph{k} times and in each iteration, a different group is picked to be used as a validation set.  

During the training phase, we experimented with different parameters in the models in order to achieve the highest scores. These parameters are called ``hyper-parameters'' in machine learning and are not directly learned within the training. For instance, the parameters \textit{alpha}, \textit{gamma} and \textit{kernels} are passed as arguments to the desired model to be trained.

Common techniques such as GridSearch or RandomSearch check the full space of available values for the hyper-parameters, which can be time and computationally demanding. In order to more quickly explore the search space and find the most suitable combination of those parameters for each model, we applied a sequential model-based optimization~\cite{skop}. The method applies a Bayesian optimization technique that takes into account the information of previous trials to chose the next set of values. The parameters we iterated over for each model are listed in Table~\ref{tab:table_hyperparmeters}.

\begin{table}[ht]
	\centering
	\caption{Tuned hyper-parameters for each model.}
	\label{tab:table_hyperparmeters}
	\begin{tabular}
		{p{2.0cm}p{6cm}}
		\toprule
		Models & Parameters \\
		\midrule
		
		\multirow[t]{1}{*}{Elastic Net}  	&	alpha, l1\_ratio, tol, max\_iter    \\[1mm]
		\multirow[t]{2}{*}{Random Forest}  & max\_features, min\_samples\_split, \\
										& bootstrap, n\_estimators \\[1mm]
		\multirow[t]{2}{*}{SVM}  	  		& kernel, C, tol, coef0, degree, gamma, \\
										& epsilon \\[1mm]
		\multirow[t]{3}{*}{MLP}  			& hidden\_layer\_sizes, alpha activation,\\
										& learning\_rate, learning\_rate\_init, tol \\
		\bottomrule
	\end{tabular}
\end{table}

We show the accuracy by using the coefficient of determination, $R$-squared ($R^2$). 
For a given prediction function $y=f(x)$, $R^2$ determines how much the total variation of $Y$ (dependent variable) is due to $X$ (independent variable). In other words, it is $1\, - Z$, where $Z$ is the ratio of the residual sum of squares to the total sum of squares, as in Equation (\ref{eq:r2}):
\begin{equation}
    R\textsuperscript{2} = 1 - \dfrac{\sum(Y_{actual}-Y_{predicted})^{2}}{\sum(Y_{actual}-Y_{mean})^{2}}
    \label{eq:r2}
\end{equation}
\vspace{2mm}

We applied $R^2$ as measure of validity because it is widely used indicator in machine learning and statistical analysis.

\subsection{Input Feature Selection}\label{sec:feat-sel}

In this step, we want to simplify our methodology by identifying the input features that will most affect the prediction, while reducing the time required to collect the PMC profiles from the applications when building new training sets. To have a common ground across multiple architectures, we select a generic PMC subset (found in all modern architectures) when building our prediction models. By using only generic features, we will build a simple and a more portable methodology to be applied on several architectures without depending on specific PMC hardware support.

Our feature selection process is quite simple. First, we looked at the list of generic counters provided by the standard profiling tool on Linux (the \verb@perf@ tool). We selected a subset of those counters that allowed us to collect all of them without requiring (1) to perform PMC multiplexing on the limited physical PMC registers and (2) to run the applications several times with different PMCs due to conflicts in simultaneously collecting PMCs. From the \verb@perf@ tool list, we selected the counters presented in Table~\ref{tab:tab_select_counters}. 

\begin{table}[t]
	\centering
	\caption{Selected hardware counters and derived metrics for feature selection step.}
	\label{tab:tab_select_counters}
    \begin{tabular}
    {
    p{1.3cm}p{6.5cm}}
    \toprule
    
    Events & Counters \\
    \midrule
	
    \multirow[t]{2}{*}{Hardware}  & 	cycles, instructions, branch\_instructions\\ &  branch\_misses \\[1mm]
    \multirow[t]{1}{*}{Software}  	  & page\_faults,  context\_switches, cpu\_migrations \\[1mm]
     \multirow[t]{1}{*}{Cache}  	  & cache\_references, cache\_misses\\[1mm]
     \multirow[t]{1}{*}{\emph{Calculated}}  & \emph{CPU\_usage} \\ 
	\bottomrule
    \end{tabular}
\end{table}

In Section \ref{res:offline}, we evaluate the quality of the schedule via simulation using \emph{generic PMCs} compared to using \emph{all counters} originally used in the model training (given in Table \ref{tab:table_metrics}).

\begin{figure*}[ht]
	\begin{center}
		\includegraphics[scale=0.65]{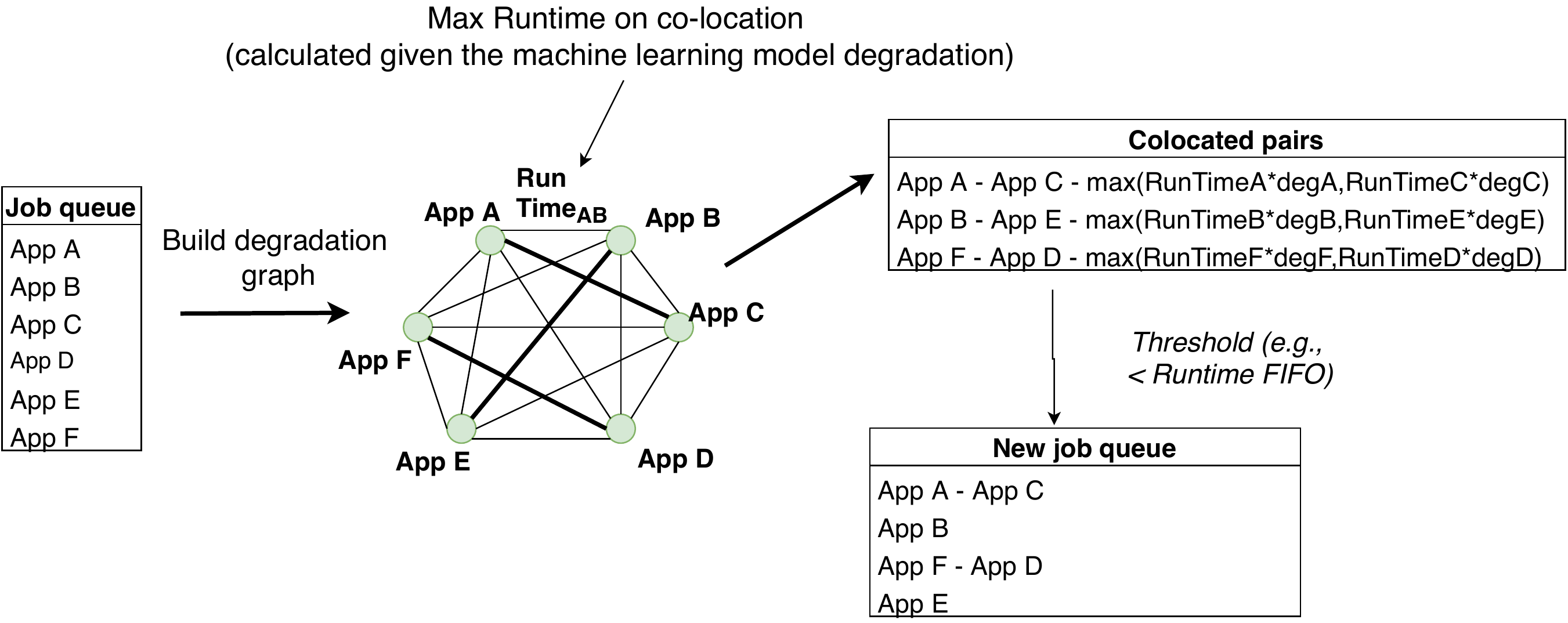}
		\caption{Overview of the scheduling phase using a degradation graph built using predictions from the selected machine learning model.}
		\label{fig:sched}
	\end{center}
\end{figure*}

\subsection{Model Deployment}\label{sec:deploy}

After training and validating the models, we generate and schedule randomized queues of applications with the goal of minimizing the overall execution time. Given a set of \textit{n} independent applications $A_1, A_2, ..., A_n$, the goal is find a schedule that  colocates a pair of applications ($i$ and $j$) onto a server node with minimum execution time, expressed by the term $\max ( RunTime_{i}*Deg_{i,j}, RunTime_{j}*Deg_{j,i} )$. The total cost of the schedule to be minimized is the sum of all these terms for the chosen pairs.

For two applications sharing the same node, given the degraded execution time as an input, the optimal co-scheduling can be found in polynomial time~\cite{jiang2008analysis}. 
This problem can be modeled as a fully connected graph, the \emph{degradation graph}. As shown in Figure~\ref{fig:sched}, each vertex of the degradation graph is an application (from the job queue) and each edge is the highest degraded runtime, calculated using the predicted degradation of the pair when they run on the same node. 

The optimal colocation problem for two applications can be viewed as a \textit{minimum-weight perfect matching} problem given the degradation graph{~\cite{jiang2008analysis}}. 
Finding the solution for this matching problem is equivalent to finding an optimal colocation scheduling for pairs of applications because a valid scheduling means all selected pairs should cover all vertices without sharing the same vertex, which is a condition for a perfect matching. The minimum-weight ensures the objective function of the colocation problem is satisfied, which stands for minimizing the sum of the execution times, taking into account the degradation between colocated applications. 

We use the \textit{blossom} algorithm~\cite{blossom} to optimally solve the matching problem for application pairs in polynomial-time: $\bigO(n^2m)$, where $n$ and $m$ are the number of nodes and edges in the degradation graph, respectively.  
We show in Section~\ref{subsec:time} an analysis on the computational time required to solve this matching problem and produce a scheduling solution, based on the degradation graph and approximately using a greedy heuristic, which simply selects the pairs with the lowest runtime to execute once the pairs are ready to be scheduled.

While the output given by the solution of the \textit{minimum-weight perfect matching} results in a set with minimum overall execution time for the entire set of pairs, the solution may produce pairs with very high execution time due to degradation when compared to its solo/serial execution. Thus, we decide to schedule those pairs with excessive degraded runtime in a serial fashion (using exclusive resources). This is because excessive degradation will increase the \emph{makespan} when executing those pair of applications together in the system.

\subsubsection{Simulated Scenario}\label{sec:sim-deploy}

Before deploying the models on the real system, we perform an offline deployment on which we project the performance of the models while generating scheduling for different queues without executing them in the real server. The simulation gives us a good approximation of the real execution as we compute the performance of the schedule using the real measures previously collected for a given set of applications. Thus, we can use the real degradation and runtime of the applications to compute the projected \emph{makespan} of each model solution. This allows us to perform a more accurate comparison across the models beyond the trained model indicator (R2 score). After performing the simulations, we select the model that most minimized the \emph{makespan} to be deployed and validated upon execution on a real system scenario. In Section \ref{sec:real_deployment}, we show the numbers for the model results.

\subsubsection{Real System}\label{sec:real-deploy}

For the final deployment on the real system, we incorporate the trained and simulated model into an existing batch scheduling system. We implemented and deployed the chosen models in the Slurm workload manager. We have
adopted Slurm~\cite{yoo2003slurm} because it is a popular, scalable, widely deployed, open-source, fault-tolerant job scheduling system for Linux clusters. Our solution is implemented as a scheduling plugin that provides
the necessary structure to colocate the jobs. The plugin is executed periodically to (1) check the queue of submitted jobs and (2) compute the pair of jobs that are allowed to run together. After computing the schedule, the
plugin attempts to find colocation opportunities for the pending jobs on the server machines that are already running other jobs.

\section{Experimental Evaluation}
\label{sec:results}

We evaluate our approach in a cluster of multi-core servers running several HPC applications. We perform several experiments with varying job queues to compare our approach with two existing default policies in Slurm's workload manager: (a) the first executes the applications sequentially on a server, and (b) the second allows for server sharing, but without any degradation knowledge. We also compare (c) a greedy solution vs blossom (perfect matching) solution given the knowledge of the degraded runtime execution graph. The goal is to contrast the quality of output (metric of interest is makespan) and time required to produce a scheduling solution.

\begin{table}[t]
\caption{Applications used in our experiments.}
    \begin{tabular}{@{}ll@{}}
		\toprule
		\textbf{Benchmark } &  \textbf{Applications} 					    \\
		\textbf{Suite}      &       \\
		\midrule
		Parsec  	    &  \textit{blackscholes, fluidanimate, swaptions, } \\
		                &  \textit{vips, streamcluster    }        \\[1mm] 
		Rodinia 	    &  \textit{lavaMD, lud, cfd, }
		particlefilter            				    \\[1mm] 
		MineBench    	&  \textit{hop, SVM-RFE, kmeans      				}				 \\[1mm] 
		NPB			    &  \textit{bt, ft, lu, sp, ua, ep, cg         		}					  \\ [1mm]
		Splash 		    &  \textit{barnes, fmm, ocean, waternsquared}         							  \\ 
		                & \textit{waterspatial} \\[1mm]
		Miscellaneous   &  \textit{stream, lulesh, SSCA, fft, mandelbrot, qsort,}\\
		                & \textit{miniFE, HPCCG} \\         
		\bottomrule
	\end{tabular}
		\centering
	
	\label{tab:table_benchmark}
\end{table}

\begin{table*}[t]
	\caption{Parameters tested during training phase.}
	\resizebox{\textwidth}{!}{ 
		\begin{tabular}{@{}l>{\raggedright\let\newline\\\arraybackslash\hspace{0pt}}p{5.5cm}lll@{}}
			\toprule
			\textbf{Models}               		  & \textbf{Hyper-Parameters}              	   & \textbf{Counters}          & \textbf{Best values --- mean/min/max/sd} & \textbf{Best values --- mean} \\ \midrule
			\multirow[t]{2}{*}{\textit{Elastic Net}} & \multirow{1}{*}{alpha, l1\_ratio, tol, max\_iter}   &
			All      &    1, 1.0, 1.0e-05, 10000     &    0, 1.0, 0.001, 10000     \\[0.7mm]
			&                        & Selected  &   0, 1.0, 0.001, 10000       &     0, 0.048, 1.130e-05, 1240     \\[1mm]
			\multirow[t]{2}{*}{\textit{SVM}} 		  & \multirow[t]{2}{5.5cm}{kernel, C, tol, coef0, degree, gamma, \\ epsilon}
			&\multirow{1}{*}{}All   &    'linear', 100.0, 0.001, 1.0, 1, 'scale', 100.0  & 'linear', 100.0, 1.0e-05, 1.0, 5, 'auto', 100.0   \\[0.7mm]
			&       
			&\multirow{2}{*}{}Selected  &    'poly', 52.260, 1.0e-05, 0.966, 3, 'auto', 100.0       &     'poly', 100.0, 1.0e-05, 1.0, 3, 'auto', 100.0     \\[1mm]
			\multirow[t]{2}{*}{\textit{Random Forest}} & \multirow[t]{1}{5.5cm}{max\_features, min\_samples\_split, bootstrap, n\_estimators}   & All      &  'auto', 7, True, 10       &   'auto', 3, True, 9      \\[0.7mm]
			&                          	   &Selected  &  'auto', 7, True, 15      &     'sqrt', 2, True, 22       \\[1mm]
			\multirow[t]{2}{*}{\textit{MLP Regressor}} & \multirow[t]{2}{5.5cm}{hidden\_layer\_sizes, alpha, activation, learning\_rate, learning\_rate\_init, tol} &\multirow{2}{*}{}All      &   (32,32,8), 0.0898, relu, adaptive, 0.0002, 0.003 &     (256,256), 0.0003, identity, adaptive, 0.0001, 1.0      \\[0.7mm]
			&      &\multirow{1}{*}{}Selected  &     (128,128), 0.985, logistic, invscaling, 0.532, 0.007   &  (32,256,256), 0.0270, logistic, invscaling, 1.0, 0.0001    \\
			\bottomrule   
		\end{tabular}
	}
	\centering
	\label{tab:param_training}
\end{table*}

\subsection{Experimental Setup}\label{sec:setup}

We carried out the experiments on a cluster set up with servers equipped with two Intel Xeon SandyBridge-EP E5-2670 that together comprise 16 cores operating at 2.6 GHz. Each socket has 20MB L3 cache(LLC) shared among all cores. Each server has 64 GB of DDR3-1600 DIMMs main memory. 

To minimize Linux interference on the makespan, the CPU \textit{intel\_pstate} governor was set to ``performance'' mode with the clock speed fixed at the highest frequency. The operational system was SUSE Linux Enterprise Server 11 SP3 with kernel version 3.0.101-0.47.90-default x86\_64. For our experiments, we considered a set up of a head node that executes the Slurm controller daemon and the compute nodes up to 5 servers. Hyperthreading was disabled as in most data center systems.

We used a total of 32 applications from different benchmarks to perform our experiments, including 27 applications from Parsec~\cite{Bienia:2008:PBS:1454115.1454128}, Rodinia~\cite{Che:2009:RBS:1678998.1680782}, NPB~\cite{Bailey:1991:NPB:125826.125925} and 5 applications from Splash~\cite{sakalis2016splash}. The applications on Table~\ref{tab:table_benchmark} were selected to cover a variety of computational patterns found in multithreaded and high-performance codes. All applications were compiled using GNU/Linux GCC 6.2.0 with \verb:O3: optimization flag and multithreading enabled, so each application could use the number of threads equals the number of cores on the server node. We profiled the application using the Linux tool \texttt{Perf} at a per-thread level and the working set for the applications was tuned to exceed the size of private caches, as it is common for native input size on real machines.

We used the \textit{Scikit-learn} library~\cite{sckitlearn} to implement the machine learning techniques. 
The accuracy of the models is calculated using the coefficient of determination, $R$-squared ($R^2$), as explained in Section~\ref{subsec:r2}. Recall that the performance of the model is better when the value of the $R^2$ is close to one. 

\subsection{Accuracy of the different models}\label{res:model_eval}

\begin{figure}[t]
	\begin{center}
		\includegraphics[scale=0.175]{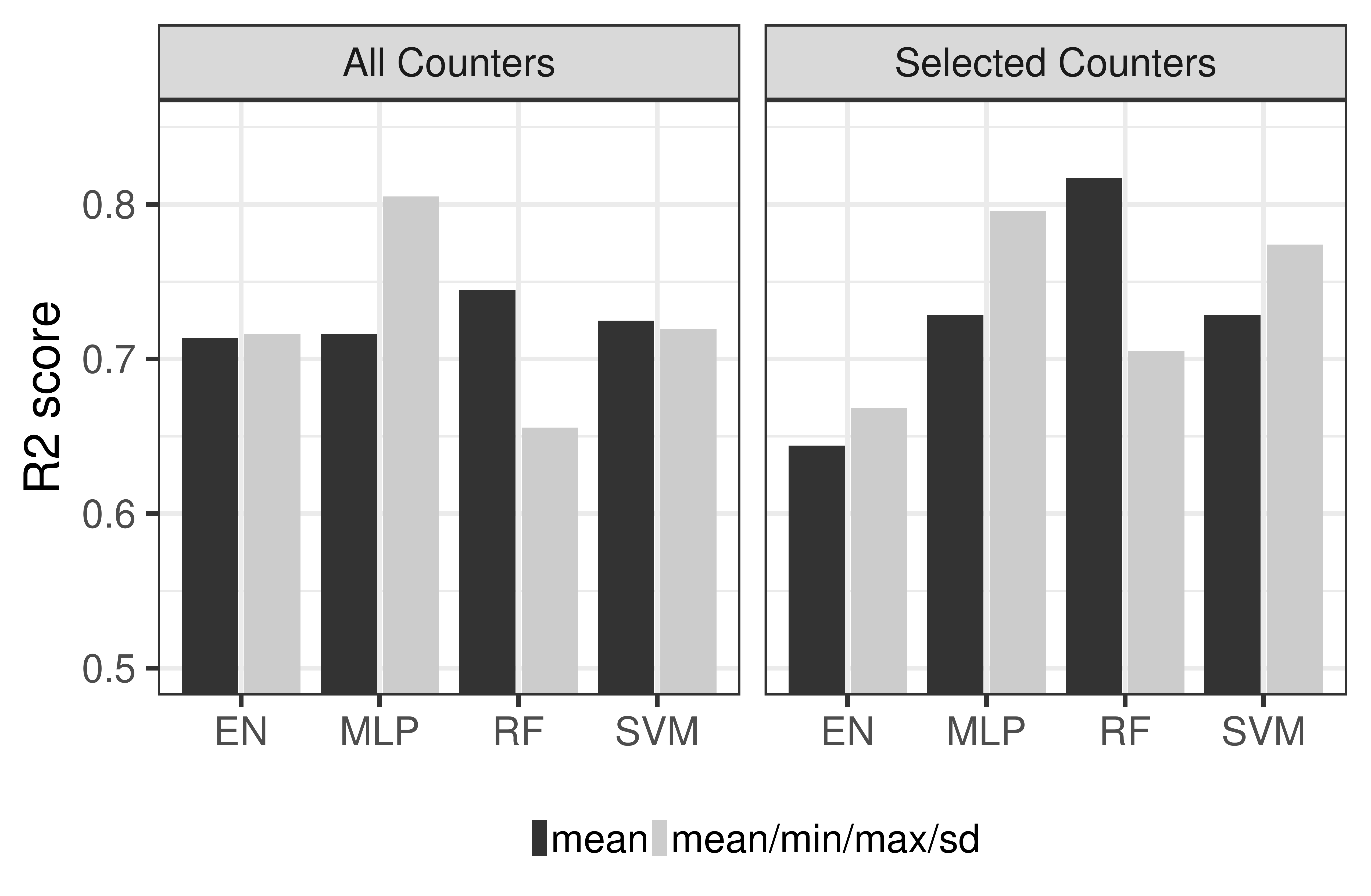}
		\caption{Highest accuracy achieved using \emph{all collected counters} and  the subset of  \emph{selected counters}.}
		\label{fig:acurracy_all}
	\end{center}
\end{figure}

To analyze the accuracy of the different models, we first present the training and validation results following the steps described in Section~\ref{sec:approach}. During the training phase, we also evaluated the impact of using additional information (such as min, max, and standard deviation) for each collected PMC feature, as well as decreasing the number of required features for prediction. We then built and trained different models based on each scenario. Table~\ref{tab:param_training} presents additional information about the hyper-parameters selected for each model used in this section. Figure~\ref{fig:acurracy_all} shows the highest accuracy (\textit{r2\_score}) for the trained models in our validation step. For brevity and clarity on the figures we will refer to the models as following: \emph{EN} for Elastic Net, \emph{SVM} for Support Vector Machine, \emph{RF} for Random Forest and \emph{MLP} for Multilayer Perceptron. 

The left-hand side of the Figure~\ref{fig:acurracy_all} shows the highest accuracy  reached by each model when using all collected hardware counters (Section~\ref{sec:data-collect}). We can observe that all machine learning techniques achieved an accuracy of about 70\%. We also noticed that increasing the feature space with \emph{min}, \emph{max} and \emph{standard deviation} values does not drastically improve the accuracy for ElasticNet and SVM models. On the other hand, it had the opposite effect for MLP and Random Forest accuracy. The accuracy for MLP increased 9\%, while the accuracy for Random Forest dropped the same figure.

Since we decided to simplify the methodology by using a subset of the collected PMCs as detailed in the Section~\ref{sec:feat-sel}, we trained and evaluated the models taking into account this subset of features, which are generic PMCs. The right-hand side of Figure~\ref{fig:acurracy_all} shows the accuracy of the models. We notice that reducing the number of features can decrease the accuracy for the ElasticNet model, while for SVM it has a slight improvement when compared with using all collected counters (left-hand side). The MLP model kept a similar performance and had a small improvement when using only mean values to describe its features. The Random Forest model increases the accuracy of both models with the subset of counters, reaching 80\% of accuracy. This demonstrates that using a subset of counters with a Random Forest model has the potential to perform well in the deployment. To assess its performance we carried out offline simulations in order to evaluate the quality of the schedule produced by this model.

\subsection{Makespan analysis via simulation}\label{res:offline}

\begin{figure}[t]
	\begin{center}
		\includegraphics[scale=0.175]{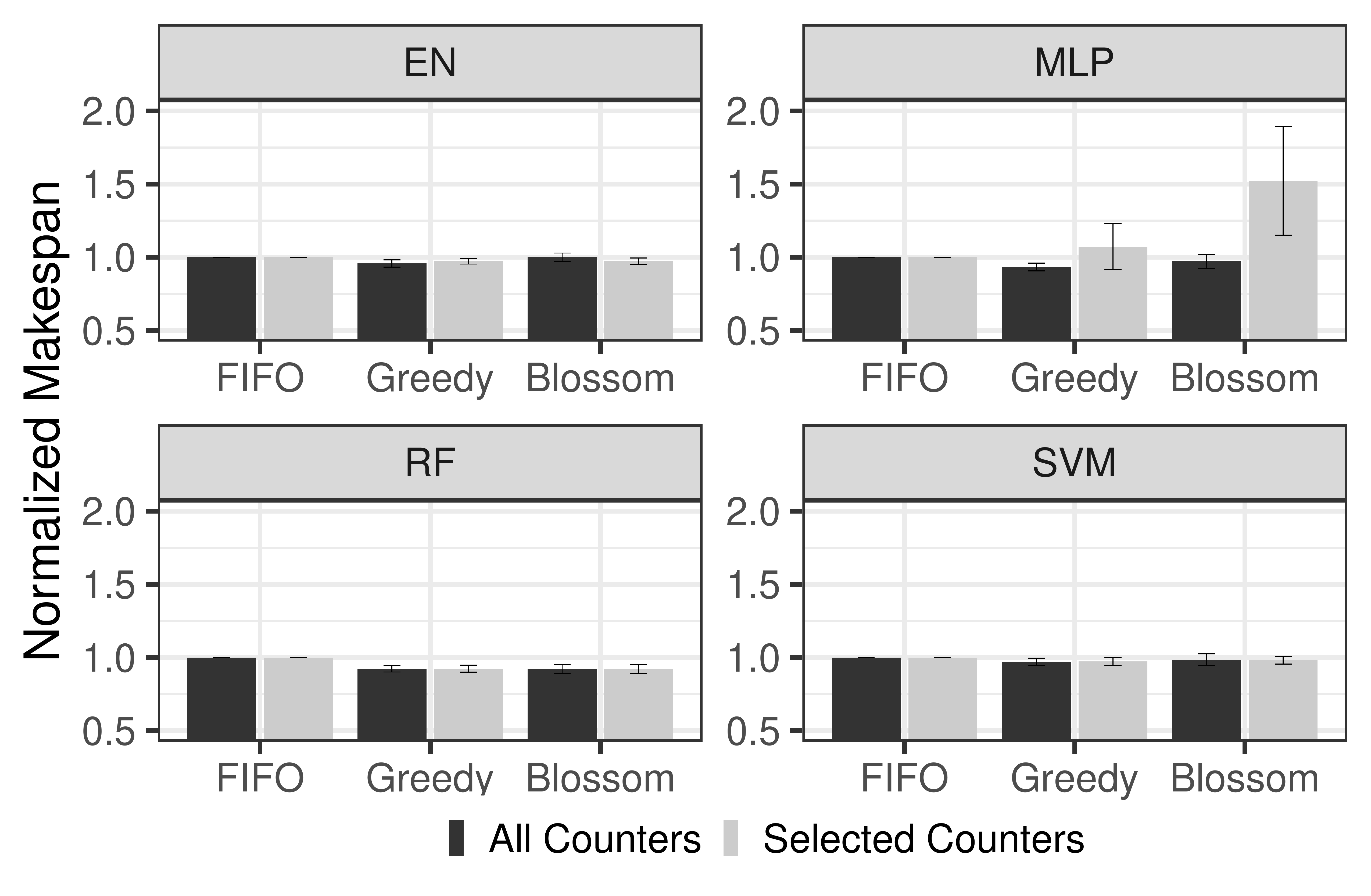}
		\caption{Performance of the models during simulated deployment. We compare the schedule pairs produced by the models compared to the FIFO approach. Each model used two different strategies to assemble the final solution: Blossom and Greedy. 
		}
		\label{fig:deployment_offline}
	\end{center}
\end{figure}

\begin{figure}[t]
	\begin{center}
		\includegraphics[scale=0.175]{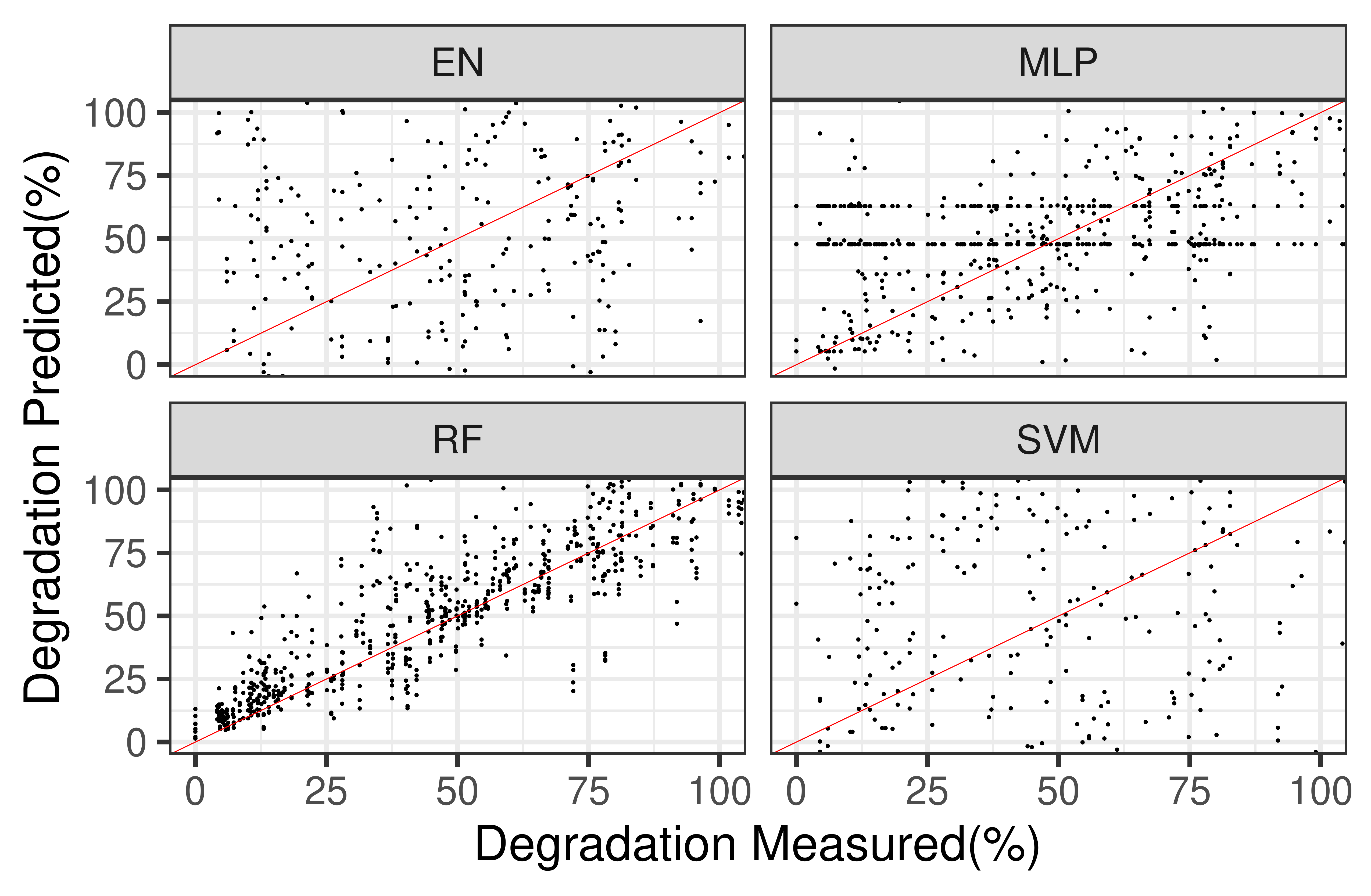}
		\caption{Prediction behavior for each model during validation with an unseen dataset.}
		\label{fig:deployment_offline_scatter}
	\end{center}
\end{figure}

The results in Section~\ref{res:model_eval} show that different models can have similar accuracy. To have a better understanding of how the models would perform in the real deployment, we simulate an offline deployment to assess the performance of the models when scheduling 20 randomized queues of submitted jobs. Rather than executing the applications in the batch scheduling system, we use the data collected for training (measured runtime for both solo and colocated settings).

The summarized simulation results for each model are shown in Figure~\ref{fig:deployment_offline}.
From the figure, we notice that for the models built with all counters, their average performance was similar to FIFO. While ElasticNet, SVM, and MLP had a slight overall improvement of less than 2\%, the RF model had an improvement of about 8\%. ElasticNet, SVM, and Random Forest showed similar performance when using the simplified/generic input features. Surprisingly, despite achieving similar high accuracy compared to the other models during the training phase, the MLP model presented a 50\% increase in the makespan when the model was simulated using the subset of features.

The figure also shows the results for two different strategies used to create the scheduling. As described in Section~\ref{sec:deploy}, the \emph{Blossom} uses the result of the matching problem over a degradation graph to provide the solution, while the \emph{Greedy} heuristic uses an ordered list to provide the solution. From the simulation, we noticed that the simplest strategy had similar performance compared to the more elaborate one. Furthermore, they kept the  same trend of performance  for every model that does not execute well in each scenario.

We further investigated why the models had poor results when they were simulated in the makespan analysis. Figure~\ref{fig:deployment_offline_scatter} shows the scatter plot of the prediction for each model when confronted with the unseen data during validation. Predictions that deviate from the true value do not contribute to maximizing server utilization, because the scheduler would make bad colocation decisions based on mispredicted values. We notice that predicting a degradation below the true value (underestimating the degradation) can be critical because the scheduler uses this indicator to colocate applications that should not be colocated together, thus making the applications experience severe performance slowdown. Predicting a degradation above the true value is also not desirable (overestimating the degradation), but it is less critical because the scheduler can assume a conservative approach and schedule the applications in a sequential (FIFO) fashion.

Looking at  Figure~\ref{fig:deployment_offline_scatter}, we can observe that ElasticNet, SVM and MLP models produce more spread out predictions around the true value (represented in the figure by the straight red line) in which it is not possible to identify any particular trend. The Random Forest model, on the other hand, produced the majority of its prediction concentrated around the straight line. Except for Random Forest, all the other models predicted negative values (below zero) when confronted with the validation data, even though we did not have negative degradation in our dataset. Thus, we deemed those models (ElasticNet, SVM, and MLP) not suitable to be used in the resource manager as they could lead to bad application pairs and an increase in the makespan to execute the queue of jobs. It was also corroborated by the results of the makespan analysis through model simulation (Figure~\ref{fig:deployment_offline}).

In conclusion, Random Forest was found to be the model with the best performance in our offline deployment. In the  next experiments, we  will show how the model accuracy influences the scheduler decisions in a real setting.

\begin{figure*}[t]
	\begin{center}
		\includegraphics[width=.87\paperwidth,height=0.27\paperheight]{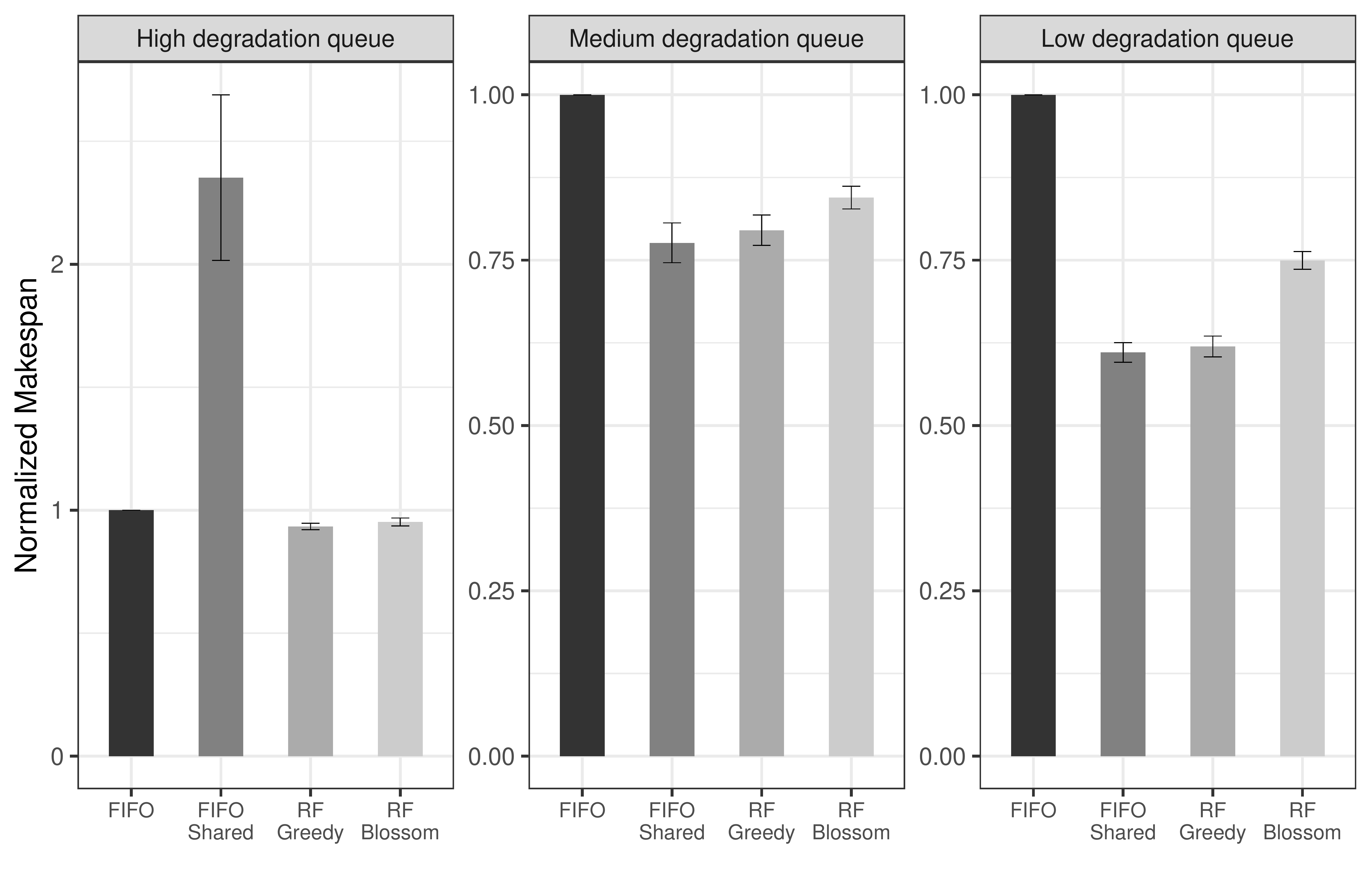}
		\caption{Normalized makespan of 5 queues with 50 applications (on a single server) for high, medium, and low degraded runtime arrival pattern. It also shows the performance of the solution using greedy and blossom strategies. }
		\label{fig:good_worst}
	\end{center}
\end{figure*}

\begin{figure}[t]
	\centering
	\begin{subfigure}[b]{0.49\textwidth}
		\includegraphics[width=1\linewidth]{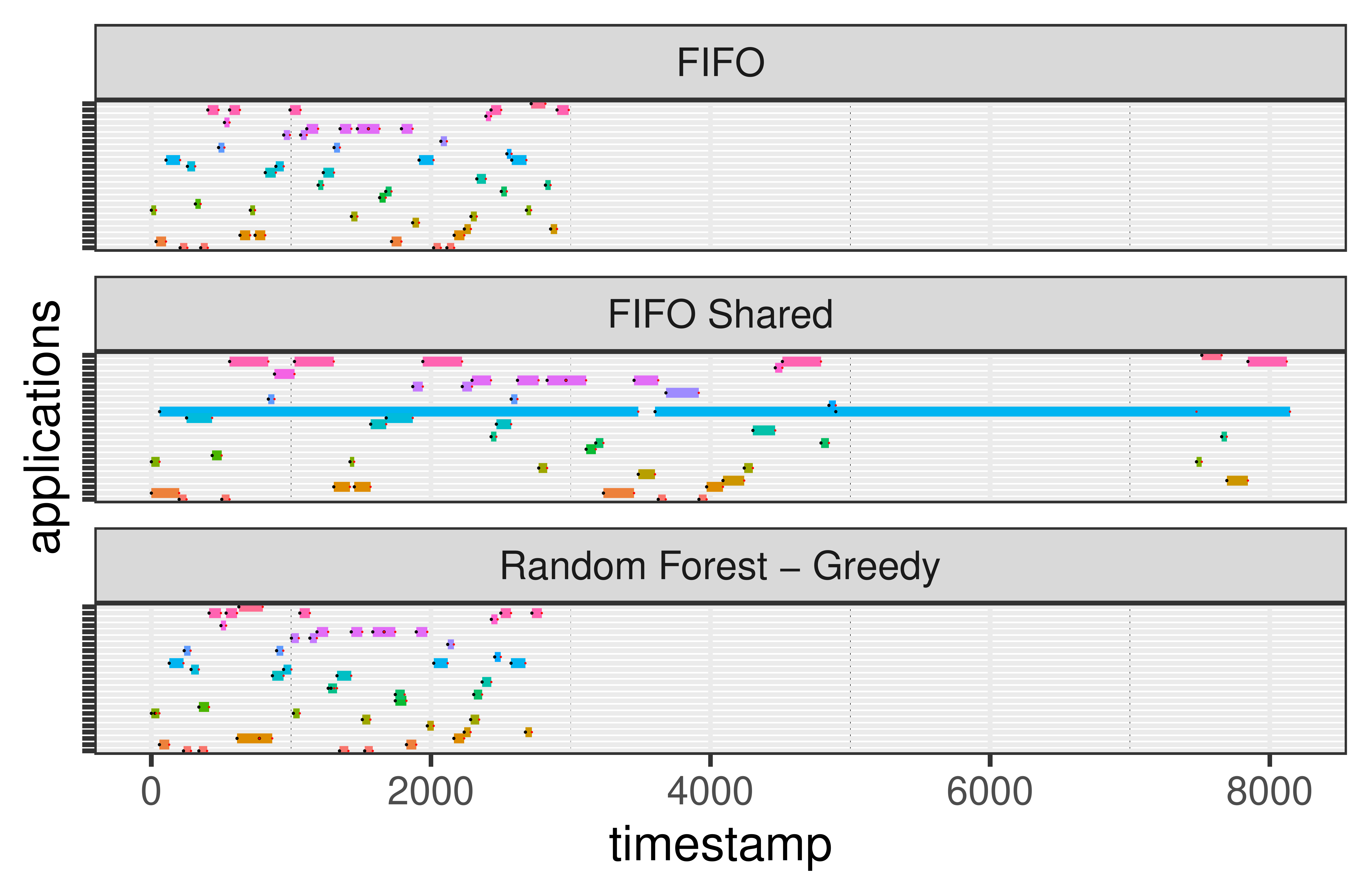}
		\caption{Job queue with high degraded runtime}
		\label{fig:high_degrad} 
	\end{subfigure}
	
	\begin{subfigure}[b]{0.49\textwidth}
		\includegraphics[width=1\linewidth]{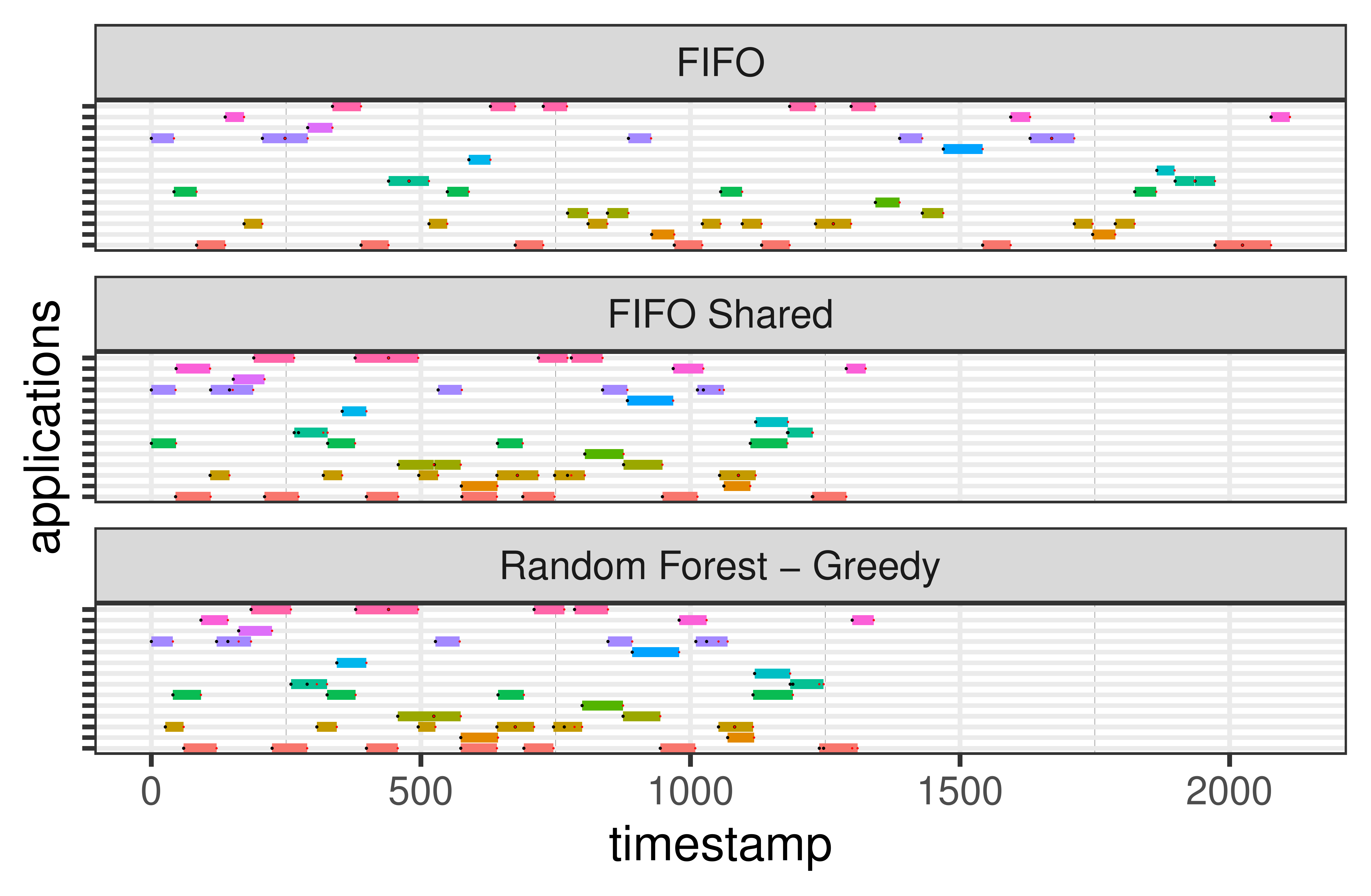}
		\caption{Job queue with low degraded runtime}
		\label{fig:low_degrad}
	\end{subfigure}
	
	\caption[High and low degradation scenarios]{
		Timeline showing how the applications are scheduled, highlighting two scenarios: (a) High degradation  vs (b) low degradation. See text for the list of applications.
		}
\end{figure}

\subsection{Real system results}\label{sec:real_deployment}

Given the promising results from model training/validation and makespan analysis performed offline, we integrate the Random Forest (best performing model) in our scheduling plugin for Slurm. As a baseline, we disable the shared resource and let the resource manager apply its FIFO execution, which we called \textit{FIFO}. Slurm allows the applications to share the whole allocated server without restrictions, this approach was called \textit{FIFO Shared}. It is agnostic to degradation between applications and executes the applications in order of arrival.

\subsubsection{Predefined Degradation Levels}\label{subsec:predef}

To investigate the effect of degradation on execution performance, we evaluate the potential of sharing resources on a set of defined groups of applications with distinct degraded runtime levels then we analyze the behavior of the scheduling strategies. We randomly separated the arriving applications into 3 different pattern queues, for each pattern we created 5 different queues with 50 applications on each. The queues are represented according to their degraded runtime amount: \textit{Low degradation queue} contained pairs of applications whose max runtime levels  were lower than 75\% of its FIFO execution;  \textit{Medium degradation queue}, contained pairs of applications whose execution order would generate runtimes of more than 75\% and less than 100\% of its FIFO execution; and \textit{High degradation queue} containing pairs applications  whose degraded runtime would be higher than its FIFO execution. This analysis aims to highlight the potential worst and best scenarios given the application set for the scheduling strategies.

Figure~\ref{fig:good_worst} presents the makespan, normalized to FIFO (serial execution). As expected, on \textit{High degradation queue}, FIFO shared had the worst performance (2.3x makespan on average) compared to FIFO. Because FIFO shared is not degradation aware, it has no flexibility to rearrange  sequential bad pairs: it executes them in order of arrival, causing an excessive amount of degradation. Although the queue are made of applications with high degradation between them, by using the machine learning models we were able to outperform the FIFO execution, mainly by its capacity of rearranging the queue to find which applications can and can not execute together. It also outperformed FIFO shared as sequential bad pairs will not share resources.

On \textit{Medium degradation queue}, FIFO shared  outperformed FIFO with an improvement of about 23\% on makespan.  Even better results were obtained on \textit{Low degradation queue} when the improvement on makespan reached around 40\%. It happens because the serial execution misses opportunities when the applications can safely share resources as they do not fully utilize the resources. In these scenarios,  FIFO shared can take advantage of the combination on applications' dispatch. As long as the order of arrival is beneficial to colocation, FIFO shared can improve the makespan over FIFO.

For \textit{Medium degradation queue} and \textit{Low degradation queue}, FIFO shared presented a slight improvement compared to our approach using the blossom strategy. This happens because FIFO shared takes a more aggressive approach pairing applications in order of arrival and using the next application in line to execute when a paired application finishes earlier. Although it is an advantage if consecutive applications to a pair can safely execute together right away, it is very dependent on the arrival order to provide good results. On the other hand, the blossom strategy executes conservatively a single pair of applications at a time and executes serially those whose runtime will exceed its FIFO execution. It means that the approach will not rely on the order of arrival and will safely execute the next applications in line only after both applications in the pair finish. On the other hand, the greedy strategy has similar performance compared to FIFO shared since it takes advantage of having a list of applications that can be safely paired.

Figures \ref{fig:high_degrad} and \ref{fig:low_degrad} show the execution of  \textit{High degradation queue} (applications in Y-axis, from top to bottom: \textit{waternsquared, swaptions, streamcluster, stream, ssca, qsort, mandel, lulesh, lu, lavamd, kmeans, hpccg, hop, ft, fluidanimate, fft, ep, cg, cfd, bt, blackscholes, barnes}) and 
\textit{Low degradation queue} (applications: \textit{waterspatial, swaptions, stream, qsort, particlefilter, minife, mandel, hpccg, hop, ft, fluidanimate, fft, ep, barnes}) over time. Each horizontal bar represents the length of the execution time of each application. We can observe that in FIFO shared, an application can be placed alongside another application at the same time. To help illustrate the application execution, red and black dots in the beginning and end of each bar represent the start and end of the execution.

In the high degradation scenario, it is considerably better to serialize the execution of the applications (Slurm's FIFO policy) rather than to allow server sharing (Slurm's sharing policy). On the other hand, in the low degradation scenario, server sharing can enhance performance compared to running the applications serially because the applications are not contenting for the same shared resources. Since our approach attained good makespan reduction compared to FIFO, this result demonstrates the effectiveness of our predictive approach that could correctly identify the scenarios where it is worth performing workload colocation vs\ non-colocation.

\subsubsection{Random job queues}\label{subsec:random}

\begin{figure}[!htbp]
	\begin{center}
		\includegraphics[scale=0.17]{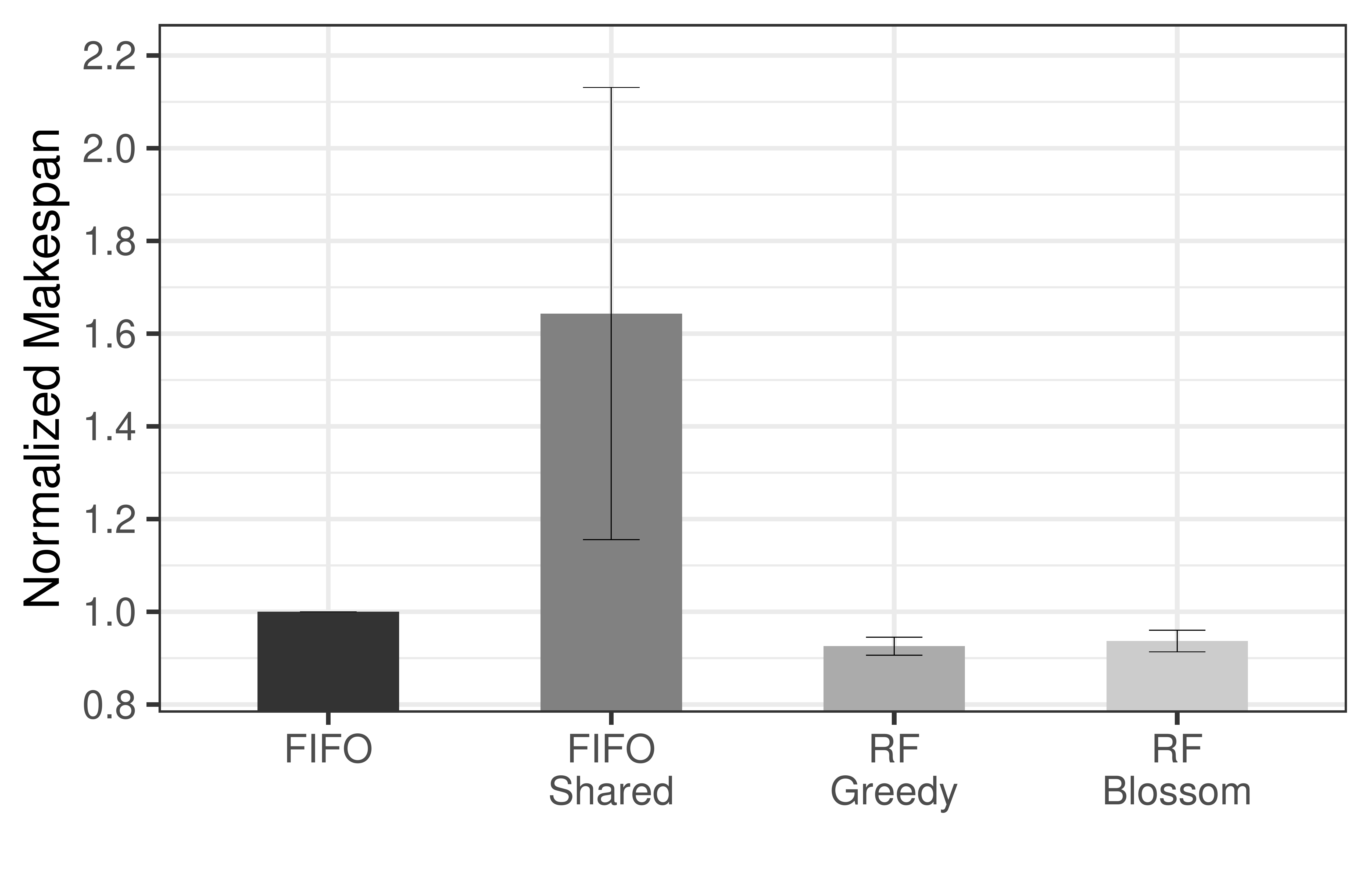}
		\caption{Normalized makespan for greedy and blossom strategies to execute 20 randomly-generated queues on a single server.}
		\label{fig:makespan_one_node}
	\end{center}
\end{figure}

For the evaluation of the  model deployed on a real system, we generated 20 queues each one containing 50 randomly chosen applications. Then we ran those queues on a single server and the batch scheduler system analyses all jobs submitted to the queue.  Figure~\ref{fig:makespan_one_node} presents the average makespan normalized to FIFO for the methodology. We note that FIFO shared increased the makespan by almost 50\%, while our solution improved the makespan on average by about 7\%. It is worth mentioning that our solution had better performance than FIFO for every executed queue. When investigating the results of the individual queues, in comparison to FIFO, our solution had 3\% of makespan improvement in the worst scenario and almost 12\% in the best scenario. {In addition, our approach also outperformed the DI heuristic\footnote{In our implementation of DI, we took advantage of the execution profile of the jobs available to the resource manager (in our case, Slurm) and sorted the jobs in ascending order list based on their LLC miss rates. Then, we build the scheduling pairs grouping an application with the lowest and another with the highest cache miss in the list until we considered all jobs. This maximizes resource usage variance to avoid shared resource contention.} presented in  Figure~\ref{fig:di_x_fifo}, which allowed bad pairs to execute together in most executed queues, thus increasing the makespan by 7\%. This result highlights the importance of considering additional performance counters when characterizing the degradation instead of relying on cache misses as a sole indicator.}

\begin{figure}[htp]
	\begin{center}
		\includegraphics[scale=0.17]{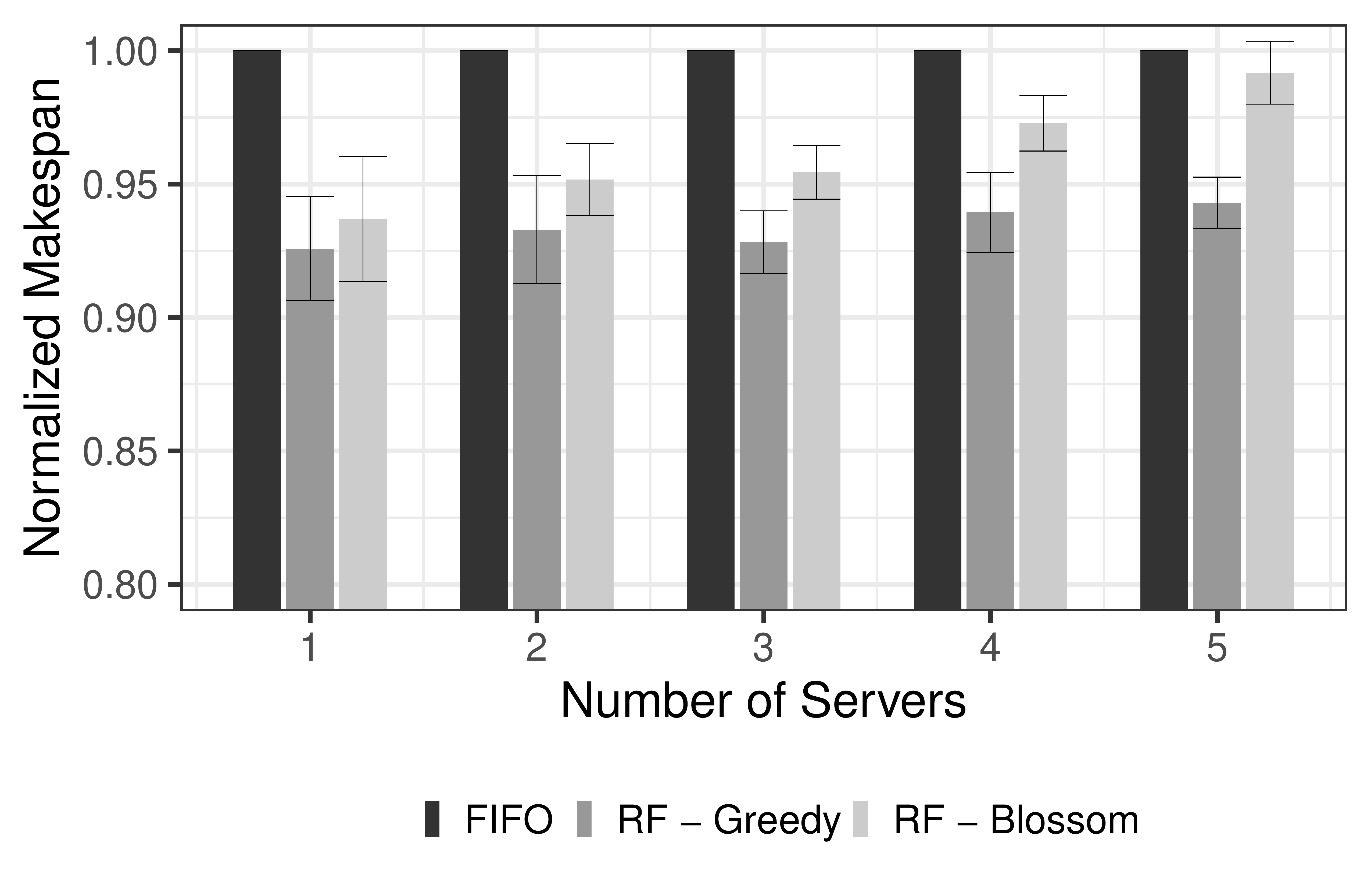}
		\caption{Normalized makespan for blossom and greedy strategies to execute 20 randomly generated queues in multiple servers.}
		\label{fig:makespan_tot_multi_node}
	\end{center}
\end{figure}

Following the good results achieved after deploying the model in a real system, we evaluated its scalability increasing the number of servers in the configuration. For each server added to the evaluation, we also increased linearly the number of dispatched jobs for each queue. Following the test for one server, in this test the batch scheduler system analyses all jobs submitted to the queue. Figure~\ref{fig:makespan_tot_multi_node} presents the makespan for each configuration when the number of servers varies. We noticed that for one to three servers the average improvement is about 5\% to 7\%, however, we had a slight decrease when the number of servers was four and five using the blossom strategy. They had a respective improvement of less than 3\% and 1\% on average. Since we take a holistic view of the queue of jobs, which means that the plugin will take into account all currently jobs dispatched to the system, consequently we will see an impact in the performance when the number of considered jobs starts to grow (see more details in Section~\ref{subsec:time}).

Figure~\ref{fig:makespan_tot_multi_node} also presents the makespan results normalized to FIFO, for random forest model using the greedy strategy. Both strategies, blossom and greedy, were able to outperform the FIFO execution on average and the greedy strategy achieved a better performance than that of the blossom strategy. It kept its performance of 7\% on average for four servers with a slight decrease to 6\% for five. This is due to the overhead associated with the greedy solution which is slightest lower than the blossom (see section~\ref{subsec:time}). It is worth mentioning that the greedy strategy had better performance than FIFO for all tested queues across the servers. Furthermore, the advantage of the greedy strategy over the blossom happens as we apply the threshold that cuts down pairs that are expected to run slower than the FIFO scheme. In this case, the greedy goal of picking always the lower runtime pairs took more advantage than the well-balanced blossom strategy, since the applications in pairs with predicted excessive runtime will be executed alone when the threshold is applied. Note that using a simpler strategy already reduces the makespan significantly (about 7\% on average) over FIFO.

\subsubsection{Scalability analysis}\label{subsec:time}

We evaluate the time required to compute a scheduling solution when varying the number of applications. The solution time includes the time to predict the degradation of each pair of applications and the time to compute the co-scheduling pairs, solving the graph matching problem for blossom strategy or search in a list for greedy strategy.

\begin{figure}[t]
	\begin{center}
		\includegraphics[scale=0.17]{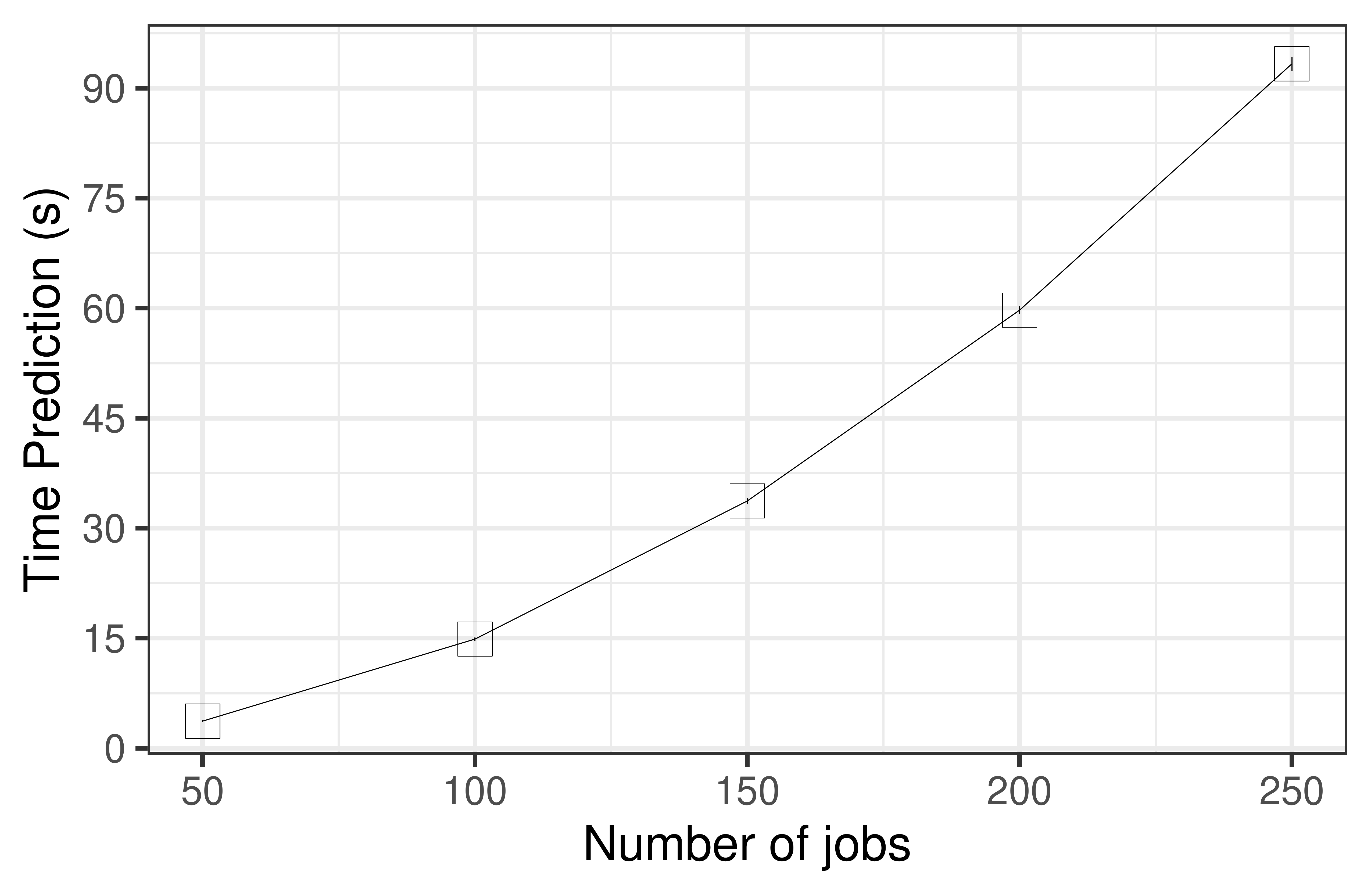}
		\caption{Measured time spent to predict the degradation for all pairs of applications using the Random Forest model}
		\label{fig:plugin_time_prediction}
	\end{center}
\end{figure}

\begin{figure}[t]
	\begin{center}
		\includegraphics[scale=0.17]{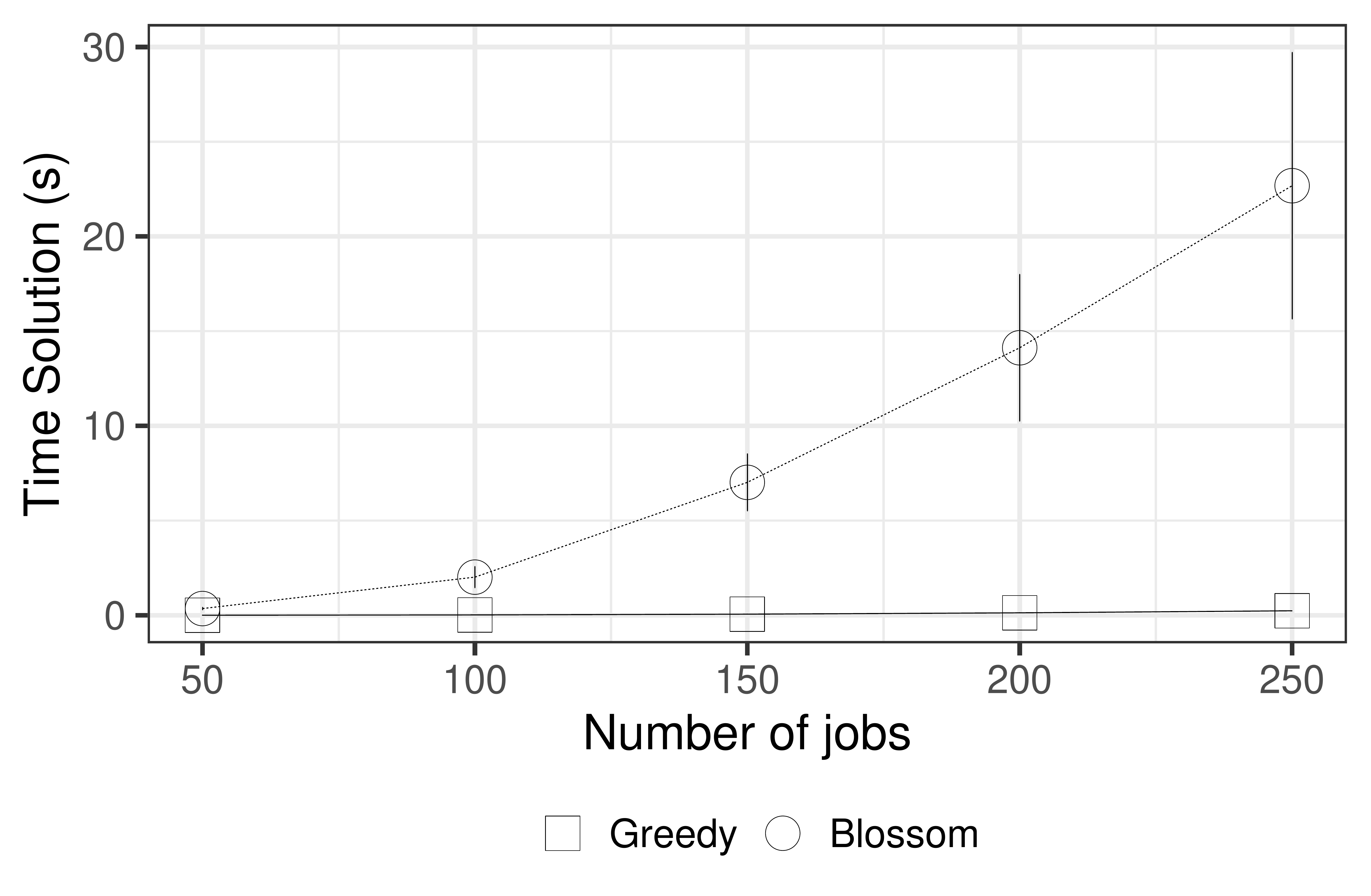}
		\caption{Measured time spent to produce scheduling decisions using two strategies: Blossom and Greedy.}
		\label{fig:plugin_time_solution}
	\end{center}
\end{figure}

Figure~\ref{fig:plugin_time_prediction} presents the time measured  to predict all degradation pairs for a varied number of jobs. This time is independent of blossom vs greedy strategies since the same model is used for both strategies. For this experiment, we found low standard deviation of 0.03\% to 0.8\%. We observe that the Random Forest model scales  almost linearly  with the number of jobs. Since the model uses as input the entire queue of pending jobs, a higher number of processed jobs can influence the model performance in the cluster scheduler.

Figure~\ref{fig:plugin_time_solution} presents the time spent creating the co-scheduling pairs, once the predicted degradation values for all pairs of jobs are generated by the model. We observe that the greedy strategy taking from 0.003~\emph{ms} to 0.23~\emph{ms} (standard deviation of less than 0.005~\emph{ms}) shows much lower processing time than blossom strategy which took from 0.35~\emph{ms} to 22~\emph{s} (standard deviation of 0.06~\emph{ms} - 7~\emph{s}). The greedy approach performs a simple sort on a list rather than executing the elaborated graph (Blossom) algorithm. The time spent to compute the scheduling for greedy never exceeded 1 second in our evaluation, while blossom spent up to 22~\emph{s}. In real settings, a greedy strategy is recommended to handle a large number of jobs in the system.

In fact, the overhead associated with the blossom strategy helps explain the  performance loss of blossom compared to greedy when adding more apps and servers in the cluster (shown in Figure \ref{fig:makespan_tot_multi_node}). For four servers the total time to create the scheduling solution was about 80~\emph{s}, while for five servers it increased to 124~\emph{s}. The greedy strategy was less affected with four and five nodes since only the prediction time dominates its overhead. On the other hand, the blossom strategy is also penalized by the time taken to create the solution that also increases with the number of jobs. Therefore, the blossom strategy will always face the highest performance penalty with a high number of jobs.

\subsubsection{Reducing model inference time}\label{subsec:reduce_overhead}

In order to improve the performance of our approach, we analyzed the impact of reducing the overhead necessary to predict the degradation for all pairs of submitted applications. As Figure~\ref{fig:plugin_time_prediction} demonstrated that the built random forest model suffers when the number of applications grows, we looked into its parameters chosen during the training phase listed in Table~\ref{tab:table_hyperparmeters} for the random forest model.

Since the random forest model averages the predictions of several base estimators, we noticed that the number of estimators is an important feature for this model and also contributes to the cost of predicting a sample. In spite of reducing the variance and increasing the accuracy of the model, sometimes a large number of estimators in a forest only increase its cost. We then decided to decrease the number of estimators evaluated for the model and training the models to find out the balance between accuracy and time to predict in order to improve its scalability.

\begin{figure}[h]
	\begin{center}
		\includegraphics[scale=0.17]{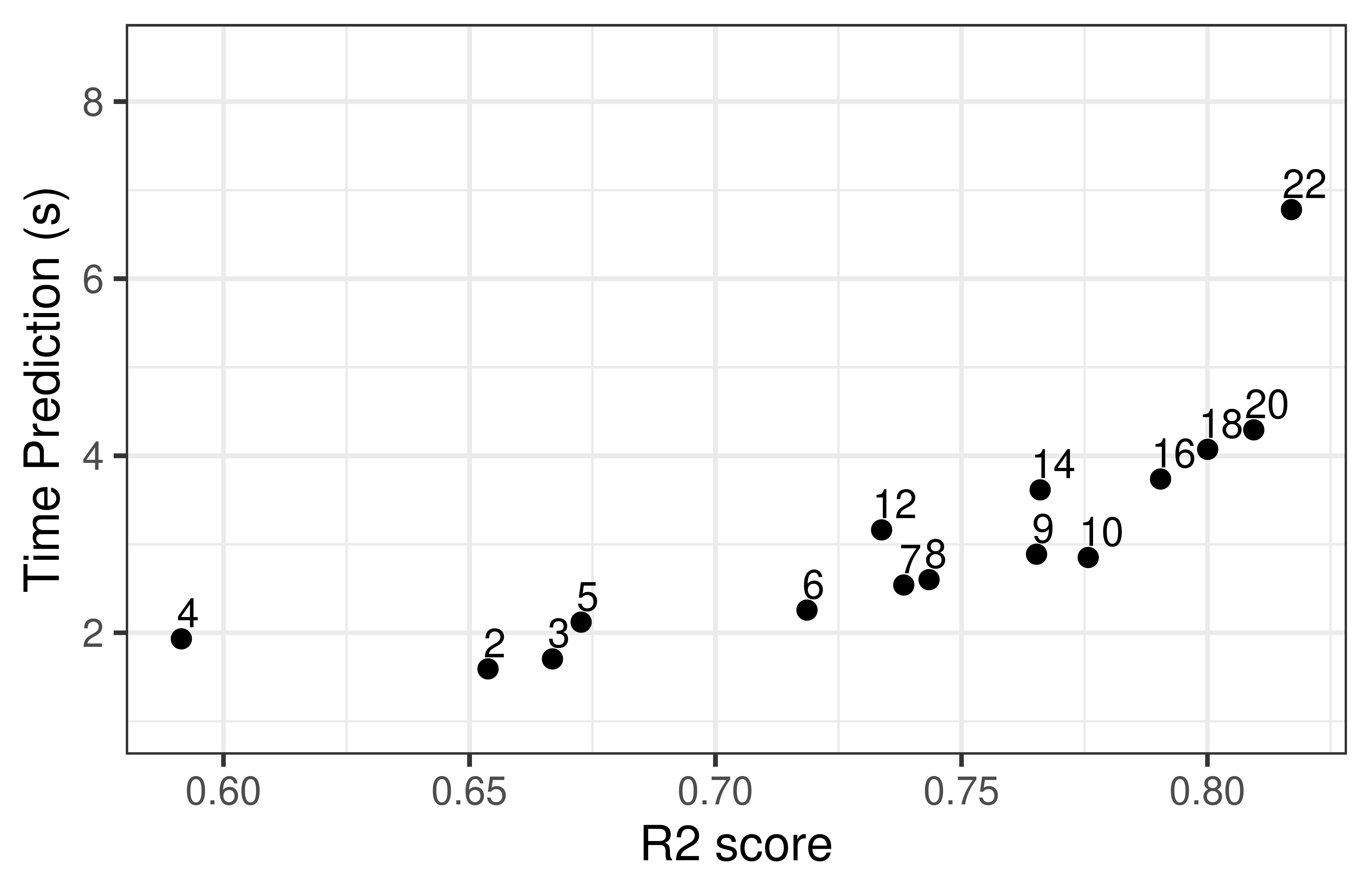}
		\caption{Time to predict the degradation for all pairs of applications for a queue of 50 applications using the random forest. Each point represents a model with a particular number of estimators. $R^2$ score is used to represent the quality of the model.}
		\label{fig:estimator_reduced}
	\end{center}
\end{figure}

Figure~\ref{fig:estimator_reduced} shows the result for the experiment of reducing the number of estimators for the random forest model. In this experiment, we trained the model with the reduced number of estimators while collected the time necessary to predict all pairs of applications for a queue of 50 applications. In the figure, we noticed that while the number of estimators increases the time taken to predict all pairs, it does not hold the same trend for accuracy. We can notice lower number of estimators with better accuracy than high number of estimators (e.g 10 and 12 estimators). In spite of using the model with the highest accuracy and also with a high number of estimators (22) for the previous experiments, we noticed the model using 6 estimators presented a good balance between accuracy and time to predict with the lowest time to predict (almost 2~s) for models with accuracy over 70\%.

\begin{figure}[h]
	\begin{center}
		\includegraphics[scale=0.17]{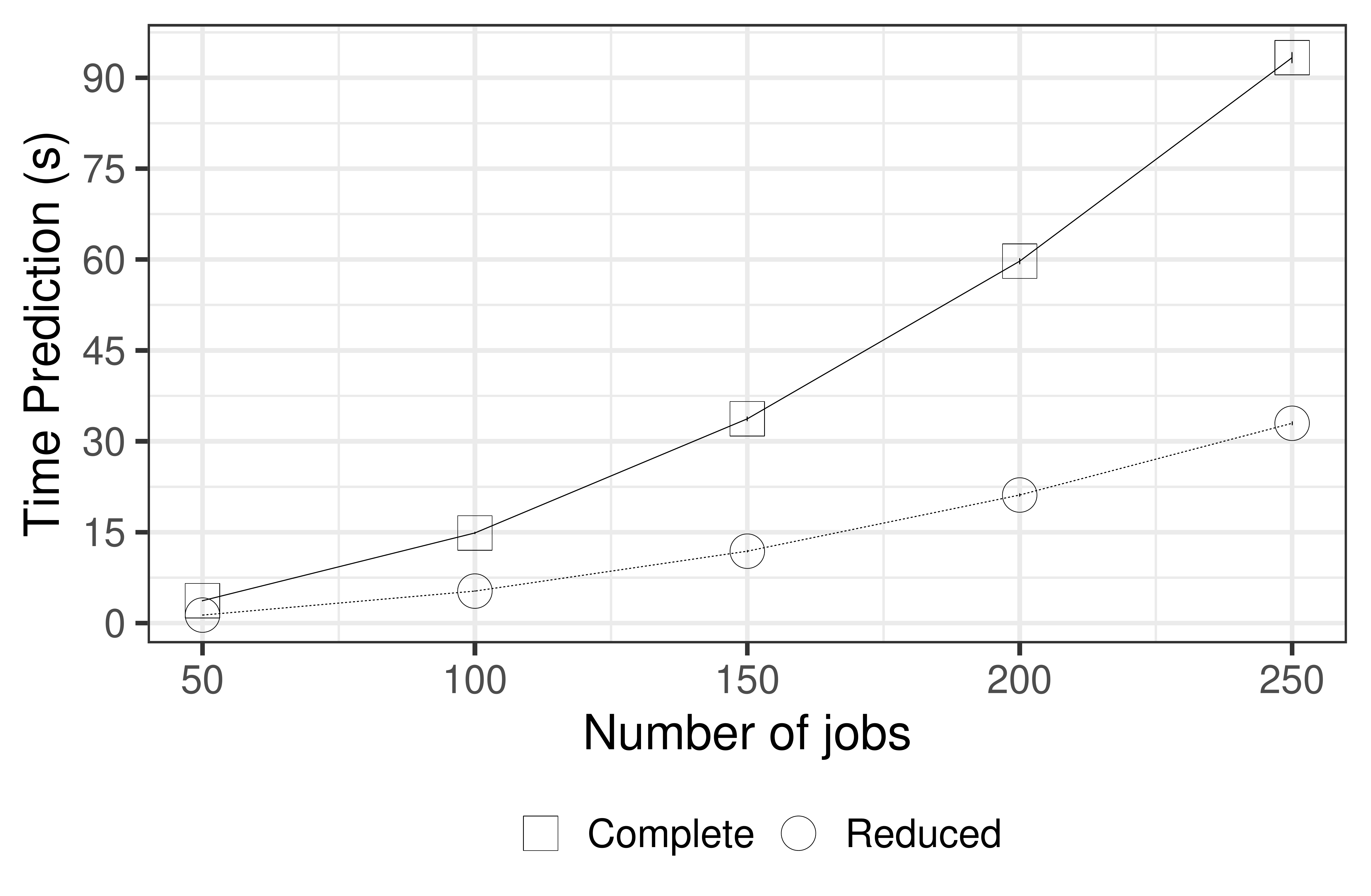}
		\caption{Measured time spent to predict the degradation for all pairs of applications using the Random Forest model from training phase with 6 estimators (Reduced) and 22 estimators (Complete).}
		\label{fig:plugin_time_prediction_reduced}
	\end{center}
\end{figure}

We compared the random forest with the reduced number of estimators against the previous model that achieved the highest accuracy while using a higher number of estimators. Figure~\ref{fig:plugin_time_prediction_reduced} shows the measured time to predict all degradation pairs using the model with the number of estimators reduced. It was measured multiple times, yielding a standard deviation of 0.01\% to 0.28\%. We clearly noticed that the reduction of the number of estimators affects the time to predict all pairs when the number of jobs increases. The time spent predicting the degradation decreased nearly 3x compared to the previous model. This decrease in the overhead will allow the approach to be used with a higher number of jobs before affecting the solution.

\begin{figure}[h]
	\begin{center}
		\includegraphics[scale=0.17]{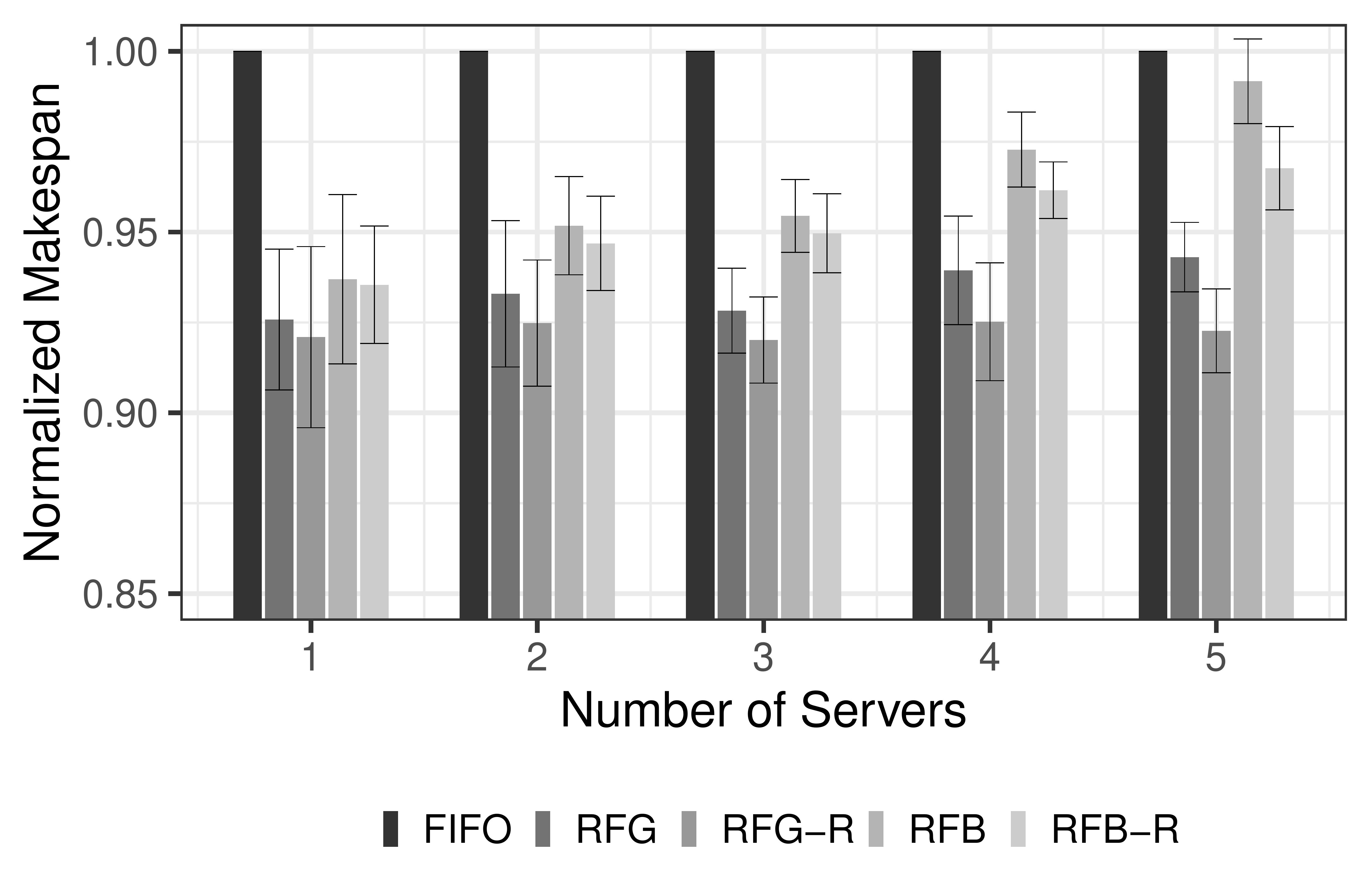}
		\caption{Normalized makespan for blossom and greedy strategies to execute 20 randomly generated queues in multiple servers. The number of estimators is reduce to decrease the random forest model's overhead (RF*-R)}
		\label{fig:rf_reduced}
	\end{center}
\end{figure}

After reducing the number of estimators for the random forest model we deployed it in the real system while varying the number of servers. Figure~\ref{fig:rf_reduced} shows the normalized makespan for each scenario. In the figure, we noticed that the reduced model using the greedy strategy (RFG-R) could be improved when the number of queued jobs increased. Since the overall time for the greedy strategy is dominated by the prediction time, it could scale better than the original model with all estimators (RFG).

On the other hand, the reduced model using the blossom strategy (RFB-R) also had a significant reduction in its overhead attaining better performance up to five servers compared to the original model (RFB). However, the blossom/optimal strategy could not scale well as the graph algorithm introduced a higher overhead with an increasing number of applications and servers in the system.

Given the results presented, we can observe that adopting a greedy strategy could best balance prediction time and effectiveness in reducing the makespan. Moreover, its lower overhead allows for scaling with a much higher number of jobs. This is important in real system deployment since our scheme could be triggered at shorter intervals without negatively impacting the scheduling manager. By tweaking key parameters in the model, we could also lower the time required to produce application degradation predictions. This was particularly important to scale our solution to a higher number of applications and servers in the system.

\section{Related Work}
\label{sec:related}

 Georgiou~\etal~\cite{georgiou2015adaptive} implemented in Slurm three powercap policies for dealing with power limitations in large-scale HPC clusters. 
Ellsworth~\etal~\cite{ellsworth2016unified} implemented power-aware Slurm plugins to explore and compare power management strategies using existing hardware platforms. 
Rajagopal~\etal~\cite{rajagopal2017novel} developed a power-aware mechanism integrated within the Slurm to improve the system resource utilization and to reduce job waiting times by redistributing the system power under
strict powercap regime. 
Sakamoto~\etal~\cite{sakamoto2017production} proposed a power-aware resource manager based on the Slurm to maintain a system-wide power limit using a set of interfaces in combination with portable APIs
for power measurement and control. Simakov~\etal~\cite{simakov2018slurm} experimented on a Slurm simulator to study the benefits of node sharing.
Dynamic Voltage and Frequency Scaling (DVFS) and Intel Running Average Power Limit (Intel RAPL) techniques have been extensively studied~\cite{6270741,sarood2014maximizing,ellsworth2015dynamic,patki2015practical} to
improve energy efficiency and system throughput under over-provisioned scenarios. 

To avoid negative interference between workloads Bubble-Up~\cite{mars2011bubble} uses the application's sensitivity and contentiousness to predict the degradation due to contention sharing the memory subsystem. The
profiling complexity for all pairwise colocations with Bubble-Up would be $\bigO(N)$ (N as the number of applications).
{Alves~\etal~\cite{alves2017multivariate} provided a model based on multiple regression analysis and trained using micro-benchmarks to predict the average slowdown of collocated applications, given the LLC, memory, and
	network pressures. The work is extended in~\cite{alves2018interference} to provide a scheduler that colocates Virtual Machines (VMs) using an Integer Linear Programming (ILP) formulation to minimize the number of physical
	machines. Octopus-Man\cite{7056037} and Hipster~\cite{7920843} allow for energy-efficient colocations of cloud workloads by exploring  heterogeneous multicore servers while focusing on a single (leaf) server.
	In contrast to these works, our solution is designed to be readily adopted as a plugin in an open-source cluster resource manager.}

Dwyer~\etal~\cite{dwyer2012practical} investigated the effectiveness of machine learning on predicting performance degradation due to colocation of applications. Their model estimates the degradation based on per-core and
system-wide hardware counter values. For multithreaded applications, the model could be used to estimate the degradation for each thread. Then, the scheduler can decide whether is necessary to allocate more hardware by
averaging the degradation of all threads.

Delimitrou~\etal~\cite{delimitrou2014quasar} proposed a cluster manager that maximizes resource utilization while meeting workload performance requirements. It uses classification to determine the
impact of the amount of resources, type of resources, and the interference on performance for each workload. The classification approach eliminates the need for exhaustive online characterization.
Regularization models~\cite{mishraesp} were previously explored for predicting application interference. These models combined linear and non-linear approaches to produce accurate predictions. The models were trained with
low-level hardware features (instructions retired, cache miss, etc.) acquired from offline measurements. 

\section{Conclusion}

We propose a solution for application colocation that predicts the degradation of colocated applications and schedules the combinations with minimum degradation. This improves the allocation of overall system computing capacity and therefore optimizes server efficiency. We experimentally demonstrate that using hardware counters in machine learning is a promising alternative to tackle this problem. We implemented our solution in the open-source cluster resource manager Slurm, as a plugin in its scheduling decision process. We carried out several experiments to evaluate the effectiveness of this method in a practical setting.

Our experimental results showed that machine learning models, in particular Random Forest, achieve good accuracy (81\%) in the evaluation phase, outperforming other models such as regression. Our solution achieved performance improvements of 7\% (avg) and 12\% (max) compared to the policies used by the Slurm resource manager. The solution based on the greedy strategy presented the best-balanced approach considering both makespan improvement and overhead to compute the scheduling solutions.

\section*{Acknowledgement}
This work is part of a project that has received funding from the European Union's Horizon 2020 research and innovation programme under Grant Agreement No. 754337 (EuroEXA); it has been supported by the Spanish Ministry of Science and Technology (project TIN2015-65316-P), Generalitat de Catalunya (contracts 2014-SGR-1051 and 2014-SGR-1272), the Severo Ochoa Programme (SEV-2015-0493) and the Ramon y Cajal fellowship (RYC2018-025628-I) of the Ministry of Economy and Competitiveness of Spain. The work is also supported by FAPESB (Edital JCB 008/2015) and the Brazilian federal government under CNPq grant (Process No. 430188/2018-8).

\bibliographystyle{elsarticle-num} 
\bibliography{ref}

\begin{thebibliography}{10}
\expandafter\ifx\csname url\endcsname\relax
  \def\url#1{\texttt{#1}}\fi
\expandafter\ifx\csname urlprefix\endcsname\relax\def\urlprefix{URL }\fi
\expandafter\ifx\csname href\endcsname\relax
  \def\href#1#2{#2} \def\path#1{#1}\fi

\bibitem{Zacarias2019colocation}
F.~V. {Zacarias}, V.~{Petrucci}, R.~{Nishtala}, P.~{Carpenter}, D.~{Moss\'{e}},
  Intelligent colocation of workloads for enhanced server efficiency, in: 2019
  31st International Symposium on Computer Architecture and High Performance
  Computing (SBAC-PAD), 2019, pp. 120--127.

\bibitem{mars2011bubble}
J.~Mars, L.~Tang, R.~Hundt, K.~Skadron, M.~L. Soffa, Bubble-up: Increasing
  utilization in modern warehouse scale computers via sensible co-locations,
  in: MICRO, ACM, 2011, pp. 248--259.

\bibitem{yang2013bubble}
H.~Yang, A.~Breslow, J.~Mars, L.~Tang, Bubble-flux: Precise online qos
  management for increased utilization in warehouse scale computers, in: ACM
  SIGARCH, Vol.~41, ACM, 2013, pp. 607--618.

\bibitem{breitbart2015case}
J.~Breitbart, J.~Weidendorfer, C.~Trinitis, Case study on co-scheduling for hpc
  applications, in: ICPPW, IEEE, 2015, pp. 277--285.

\bibitem{dutot2017towards}
P.-F. Dutot, Y.~Georgiou, D.~Glesser, L.~Lefevre, M.~Poquet, I.~Rais, Towards
  energy budget control in hpc, in: CCGrid, IEEE Press, 2017, pp. 381--390.

\bibitem{barroso2007case}
L.~A. Barroso, U.~H{\"o}lzle, The case for energy-proportional computing,
  Computer 40~(12) (2007).

\bibitem{gholkar2016power}
N.~Gholkar, F.~Mueller, B.~Rountree, Power tuning hpc jobs on power-constrained
  systems, in: PACT, ACM, 2016, pp. 179--191.

\bibitem{patki2013exploring}
T.~Patki, D.~K. Lowenthal, B.~Rountree, M.~Schulz, B.~R. De~Supinski, Exploring
  hardware overprovisioning in power-constrained, high performance computing,
  in: ICS, ACM, 2013, pp. 173--182.

\bibitem{guide2011intel}
M.~INTEL, Intel{\textregistered} 64 and ia-32 architectures software
  developer’s manual, Volume 3B: System programming Guide, Part 2 (2011).

\bibitem{zhang2016maximizing}
H.~Zhang, H.~Hoffmann, Maximizing performance under a power cap: A comparison
  of hardware, software, and hybrid techniques, ACM SIGPLAN Notices 51~(4)
  (2016) 545--559.

\bibitem{rountree2009adagio}
B.~Rountree, D.~K. Lownenthal, B.~R. De~Supinski, M.~Schulz, V.~W. Freeh,
  T.~Bletsch, Adagio: making dvs practical for complex hpc applications, in:
  ICS, ACM, 2009, pp. 460--469.

\bibitem{ellsworth2015dynamic}
D.~A. Ellsworth, A.~D. Malony, B.~Rountree, M.~Schulz, Dynamic power sharing
  for higher job throughput, in: SC, ACM, 2015, p.~80.

\bibitem{Nishtala2013emsoft}
R.~Nishtala, D.~Moss\'{e}, V.~Petrucci, Energy-aware thread co-location in
  heterogeneous multicore processors, in: EMSOFT, 2013, pp. 1--9.
\newblock \href {https://doi.org/10.1109/EMSOFT.2013.6658599}
  {\path{doi:10.1109/EMSOFT.2013.6658599}}.

\bibitem{tang2013reqos}
L.~Tang, J.~Mars, W.~Wang, T.~Dey, M.~L. Soffa, Reqos: Reactive static/dynamic
  compilation for qos in warehouse scale computers, in: ACM SIGPLAN Notices,
  Vol.~48, ACM, 2013, pp. 89--100.

\bibitem{dwyer2012practical}
T.~Dwyer, A.~Fedorova, S.~Blagodurov, M.~Roth, F.~Gaud, J.~Pei, {A practical
  method for estimating performance degradation on multicore processors, and
  its application to HPC workloads}, in: SC, IEEE, 2012, p.~83.

\bibitem{de2017disallowing}
A.~de~Blanche, T.~Lundqvist, Disallowing same-program co-schedules to improve
  efficiency in quad-core servers., in: COSH/VisorHPC@ HiPEAC, 2017, pp.
  13--20.

\bibitem{zhuravlev2012survey}
S.~Zhuravlev, J.~C. Saez, S.~Blagodurov, A.~Fedorova, M.~Prieto, Survey of
  scheduling techniques for addressing shared resources in multicore
  processors, ACM Computing Surveys (CSUR) 45~(1) (2012) 4.

\bibitem{rai2010performance}
J.~K. Rai, A.~Negi, R.~Wankar, K.~Nayak, Performance prediction on multi-core
  processors, in: CICN, IEEE, 2010, pp. 633--637.

\bibitem{subramanian2015application}
L.~Subramanian, V.~Seshadri, A.~Ghosh, S.~Khan, O.~Mutlu, The application
  slowdown model: Quantifying and controlling the impact of inter-application
  interference at shared caches and main memory, in: MICRO, IEEE, 2015, pp.
  62--75.

\bibitem{blagodurov2010contention}
S.~Blagodurov, S.~Zhuravlev, A.~Fedorova, Contention-aware scheduling on
  multicore systems, TOCS 28~(4) (2010) 8.

\bibitem{merkel2010resource}
A.~Merkel, J.~Stoess, F.~Bellosa, Resource-conscious scheduling for energy
  efficiency on multicore processors, in: EuroSys, ACM, 2010, pp. 153--166.

\bibitem{mishraesp}
N.~Mishra, J.~D. Lafferty, H.~Hoffmann, Esp: A machine learning approach to
  predicting application interference, in: 2017 IEEE International Conference
  on Autonomic Computing (ICAC), IEEE, 2017, pp. 125--134.

\bibitem{Kanev2015wsc}
S.~Kanev, J.~P. Darago, K.~Hazelwood, P.~Ranganathan, T.~Moseley, G.-Y. Wei,
  D.~Brooks, Profiling a warehouse-scale computer, in: ISCA, ACM, New York, NY,
  USA, 2015, pp. 158--169.
\newblock \href {https://doi.org/10.1145/2749469.2750392}
  {\path{doi:10.1145/2749469.2750392}}.

\bibitem{Zou05regularizationand}
H.~Zou, T.~Hastie, Regularization and variable selection via the elastic net,
  Journal of the Royal Statistical Society, Series B 67 (2005) 301--320.

\bibitem{Boser:1992:TAO:130385.130401}
B.~E. Boser, I.~M. Guyon, V.~N. Vapnik, A training algorithm for optimal margin
  classifiers, in: COLT, ACM, 1992, pp. 144--152.
\newblock \href {https://doi.org/10.1145/130385.130401}
  {\path{doi:10.1145/130385.130401}}.

\bibitem{Breiman:2001:RF:570181.570182}
L.~Breiman, \href{https://doi.org/10.1023/A:1010933404324}{Random forests},
  Mach. Learn. 45~(1) (2001) 5--32.
\newblock \href {https://doi.org/10.1023/A:1010933404324}
  {\path{doi:10.1023/A:1010933404324}}.
\newline\urlprefix\url{https://doi.org/10.1023/A:1010933404324}

\bibitem{hinton1990connectionist}
G.~E. Hinton, Connectionist learning procedures, in: Machine learning,
  Elsevier, 1990, pp. 555--610.

\bibitem{skop}
T.~Head, MechCoder, G.~Louppe, I.~Shcherbatyi, fcharras, Z.~Vinícius,
  cmmalone, C.~Schröder, nel215, N.~Campos, T.~Young, S.~Cereda, T.~Fan, rene
  rex, K.~K. Shi, J.~Schwabedal, carlosdanielcsantos, Hvass-Labs, M.~Pak,
  SoManyUsernamesTaken, F.~Callaway, L.~Estève, L.~Besson, M.~Cherti,
  K.~Pfannschmidt, F.~Linzberger, C.~Cauet, A.~Gut, A.~Mueller, A.~Fabisch,
  \href{https://doi.org/10.5281/zenodo.1207017}{scikit-optimize/scikit-optimize:
  v0.5.2} (Mar. 2018).
\newblock \href {https://doi.org/10.5281/zenodo.1207017}
  {\path{doi:10.5281/zenodo.1207017}}.
\newline\urlprefix\url{https://doi.org/10.5281/zenodo.1207017}

\bibitem{jiang2008analysis}
Y.~Jiang, X.~Shen, C.~Jie, R.~Tripathi, Analysis and approximation of optimal
  co-scheduling on chip multiprocessors, in: PACT, IEEE, 2008, pp. 220--229.

\bibitem{blossom}
J.~Edmonds, Maximum matching and a polyhedron with 0, 1-vertices, Journal of
  research of the National Bureau of Standards B 69~(125-130) (1965) 55--56.

\bibitem{yoo2003slurm}
A.~B. Yoo, M.~A. Jette, M.~Grondona, Slurm: Simple linux utility for resource
  management, in: JSSPP, Springer, 2003, pp. 44--60.

\bibitem{Bienia:2008:PBS:1454115.1454128}
C.~Bienia, S.~Kumar, J.~P. Singh, K.~Li, The parsec benchmark suite:
  Characterization and architectural implications, in: PACT, ACM, 2008, pp.
  72--81.
\newblock \href {https://doi.org/10.1145/1454115.1454128}
  {\path{doi:10.1145/1454115.1454128}}.

\bibitem{Che:2009:RBS:1678998.1680782}
S.~Che, M.~Boyer, J.~Meng, D.~Tarjan, J.~W. Sheaffer, S.-H. Lee, K.~Skadron,
  Rodinia: A benchmark suite for heterogeneous computing, in: IISWC, IEEE,
  Washington, DC, USA, 2009, pp. 44--54.
\newblock \href {https://doi.org/10.1109/IISWC.2009.5306797}
  {\path{doi:10.1109/IISWC.2009.5306797}}.

\bibitem{Bailey:1991:NPB:125826.125925}
D.~H. Bailey, E.~Barszcz, J.~T. Barton, D.~S. Browning, R.~L. Carter, L.~Dagum,
  R.~A. Fatoohi, P.~O. Frederickson, T.~A. Lasinski, R.~S. Schreiber, H.~D.
  Simon, V.~Venkatakrishnan, S.~K. Weeratunga, The nas parallel benchmarks:
  Summary and preliminary results, in: SC, ACM, New York, NY, USA, 1991, pp.
  158--165.
\newblock \href {https://doi.org/10.1145/125826.125925}
  {\path{doi:10.1145/125826.125925}}.

\bibitem{sakalis2016splash}
C.~Sakalis, C.~Leonardsson, S.~Kaxiras, A.~Ros, Splash-3: A properly
  synchronized benchmark suite for contemporary research, in: ISPASS, IEEE,
  2016, pp. 101--111.

\bibitem{sckitlearn}
F.~Pedregosa, G.~Varoquaux, A.~Gramfort, V.~Michel, B.~Thirion, O.~Grisel,
  M.~Blondel, P.~Prettenhofer, R.~Weiss, V.~Dubourg, J.~Vanderplas, A.~Passos,
  D.~Cournapeau, M.~Brucher, M.~Perrot, E.~Duchesnay, Scikit-learn: Machine
  learning in {P}ython, Journal of Machine Learning Research 12 (2011)
  2825--2830.

\bibitem{georgiou2015adaptive}
Y.~Georgiou, D.~Glesser, D.~Trystram, Adaptive resource and job management for
  limited power consumption, in: IPDPSW, IEEE, 2015, pp. 863--870.

\bibitem{ellsworth2016unified}
D.~Ellsworth, T.~Patki, M.~Schulz, B.~Rountree, A.~Malony, A unified platform
  for exploring power management strategies, in: E2SC, IEEE, 2016, pp. 24--30.

\bibitem{rajagopal2017novel}
D.~Rajagopal, D.~Tafani, Y.~Georgiou, D.~Glesser, M.~Ott, A novel approach for
  job scheduling optimizations under power cap for arm and intel hpc systems,
  in: HiPC, IEEE, 2017, pp. 142--151.

\bibitem{sakamoto2017production}
R.~Sakamoto, T.~Cao, M.~Kondo, K.~Inoue, M.~Ueda, T.~Patki, D.~Ellsworth,
  B.~Rountree, M.~Schulz, Production hardware overprovisioning: real-world
  performance optimization using an extensible power-aware resource management
  framework, in: IPDPS, IEEE, 2017, pp. 957--966.

\bibitem{simakov2018slurm}
N.~A. Simakov, R.~L. DeLeon, M.~D. Innus, M.~D. Jones, J.~P. White, S.~M.
  Gallo, A.~K. Patra, T.~R. Furlani, Slurm simulator: Improving slurm scheduler
  performance on large hpc systems by utilization of multiple controllers and
  node sharing, in: PEARC, ACM, 2018, p.~25.

\bibitem{6270741}
B.~Rountree, D.~H. Ahn, B.~R. de~Supinski, D.~K. Lowenthal, M.~Schulz, Beyond
  dvfs: A first look at performance under a hardware-enforced power bound, in:
  IPDPS, 2012, pp. 947--953.
\newblock \href {https://doi.org/10.1109/IPDPSW.2012.116}
  {\path{doi:10.1109/IPDPSW.2012.116}}.

\bibitem{sarood2014maximizing}
O.~Sarood, A.~Langer, A.~Gupta, L.~Kale, Maximizing throughput of
  overprovisioned hpc data centers under a strict power budget, in: SC, IEEE
  Press, 2014, pp. 807--818.

\bibitem{patki2015practical}
T.~Patki, D.~K. Lowenthal, A.~Sasidharan, M.~Maiterth, B.~L. Rountree,
  M.~Schulz, B.~R. De~Supinski, Practical resource management in
  power-constrained, high performance computing, in: ICPADS, ACM, 2015, pp.
  121--132.

\bibitem{alves2017multivariate}
M.~M. Alves, L.~M. de~Assump{\c{c}}{\~a}o~Drummond, A multivariate and
  quantitative model for predicting cross-application interference in virtual
  environments, Journal of Systems and Software 128 (2017) 150--163.

\bibitem{alves2018interference}
M.~M. Alves, L.~Teylo, Y.~Frota, L.~M. Drummond, An interference-aware virtual
  machine placement strategy for high performance computing applications in
  clouds, in: WSCAD, IEEE, 2018, pp. 94--100.

\bibitem{7056037}
V.~{Petrucci}, M.~A. {Laurenzano}, J.~{Doherty}, Y.~{Zhang}, D.~{Moss\'{e}},
  J.~{Mars}, L.~{Tang}, Octopus-man: Qos-driven task management for
  heterogeneous multicores in warehouse-scale computers, in: 2015 IEEE 21st
  International Symposium on High Performance Computer Architecture (HPCA),
  2015, pp. 246--258.
\newblock \href {https://doi.org/10.1109/HPCA.2015.7056037}
  {\path{doi:10.1109/HPCA.2015.7056037}}.

\bibitem{7920843}
R.~{Nishtala}, P.~{Carpenter}, V.~{Petrucci}, X.~{Martorell}, Hipster: Hybrid
  task manager for latency-critical cloud workloads, in: 2017 IEEE
  International Symposium on High Performance Computer Architecture (HPCA),
  2017, pp. 409--420.
\newblock \href {https://doi.org/10.1109/HPCA.2017.13}
  {\path{doi:10.1109/HPCA.2017.13}}.

\bibitem{delimitrou2014quasar}
C.~Delimitrou, C.~Kozyrakis, Quasar: resource-efficient and qos-aware cluster
  management, in: ACM SIGPLAN Notices, Vol.~49, ACM, 2014, pp. 127--144.

\end{thebibliography}

\end{document}